\documentclass[10pt]{revtex4}                            
\usepackage{amssymb,epsf}
\usepackage{epsfig}
\usepackage{latexsym}
\usepackage{color}

\usepackage{amsmath,amsfonts,bbm,dsfont,mathrsfs}

\usepackage{graphicx}
\usepackage[scriptsize,hang,flushleft]{caption}
\usepackage[scriptsize,hang]{subfigure}

\newcommand{\half}{\frac{1}{2}}
\newcommand{\dd}{{\mathrm{d}}}

\begin{document}
\title{Topological black holes in mimetic gravity}
\author{Ahmad Sheykhi$^{1,2}$\footnote{asheykhi@shirazu.ac.ir} and Saskia Grunau$^{2}$\footnote{saskia.grunau@uni-oldenburg.de}}
\address{$^1$Physics Department and Biruni Observatory, Shiraz
University, Shiraz 71454, Iran\\
$^2$Institut f\"{u}r Physik, Universit\"{a}t Oldenburg, Postfach
2503 D-26111 Oldenburg, Germany}

\begin{abstract}
In this paper, we construct some new classes of topological black
hole solutions in the context of mimetic gravity and investigate
their properties. We study the uncharged and charged black holes,
separately. We find the following novel results: (i) In the
absence of a potential for the mimetic field, black hole solutions
can address the flat rotation curves of spiral galaxies and
alleviate the dark matter problem without invoking particle dark
matter. Thus, mimetic gravity can provide a theoretical background
for understanding flat galactic rotation curves through modifying
Schwarzschild spacetime. (ii) We also investigate the casual
structure and physical properties of the solutions. We observe
that in the absence of a potential, our solutions are not
asymptotically flat, while in the presence of a negative constant
potential for the mimetic field, the solutions are asymptotically
anti-de Sitter (AdS). (iii) Finally, we explore the motion of
massless and massive particles and give a list of the types of
orbits. We study the differences of geodesic motion in Einstein
gravity and in mimetic gravity. In contrast to Einstein gravity,
massive particles always move on bound orbits and cannot escape
the black hole in mimetic gravity. Furthermore, we find stable
bound orbits for massless particles.

\end{abstract}

 \maketitle

\section{Introduction}
One of the most important challenges of modern cosmology is that
about 95\% of the total energy content of our Universe is dark and
the nature of these dark components is still unknown. It has been
shown that the dark side of the Universe consists of two
components called dark energy (about 69\%) and dark matter (about
26\%) of the energy content \cite{Yang}. The former has
anti-gravity property with negative pressure and is responsible
for the acceleration of the Universe expansion, and the latter has
no electromagnetic interaction but contributes to the
gravitational interaction. Indeed, it is now well established that
the baryonic matter of galaxies and clusters of galaxies does not
provide sufficient gravitation to explain the observed dynamics of
the systems and the presence of dark matter is necessary for
explanation of such dynamics. Furthermore, according to our
present understanding, dark matter has also significant
contribution to the anisotropies in the cosmic microwave
background, galaxy cluster velocity dispersions, large-scale
structure distributions, gravitational lensing investigations, and
X-ray measurements from galaxy clusters.

Of course, it is quite possible to deal with the problem in a
different perspective. Indeed, one can argue that dark matter
might be a manifestation of a theory of gravity beyond General
Relativity and does not really consists of particles. In this
regards, modified theories of gravity have been proposed to
explain gravitational lensing, flat rotation curves of galaxies
and dynamics of cluster of galaxies as a geometrical effect.
Modified Newtonian Mechanics (MOND) is a known example, which try
to explain the flat rotation curves of galaxies, through modifying
Newton's law of gravity \cite{Mil}. However, this theory suffers
to embed within a more comprehensive relativistic theory of
gravity. Another attempt for probing dark matter through a
geometrical effect is the $f(R)$ theory of gravity, which has been
investigated widely in the structure (see e.g.
\cite{Sobuti,Sobuti2,Sot,Chr,Sho,jian,Riazi,Odi,Odi2} and
references therein).

Mimetic theory of gravity was proposed a few years ago, as an
alternative description for the dark matter puzzle \cite{Mim1}. It
has been shown that the mimetic field can encodes an extra
longitudinal degree of freedom to the gravitational field. Thus,
the gravitational field achieves, in addition to two transverse
degrees of freedom, a dynamical longitudinal degree of freedom
which can play the role of mimetic dark matter. Latter, it was
shown that a modified version of mimetic gravity can resolve the
cosmological singularities \cite{MimCos1} as well as the
singularity in the center of a black hole \cite{Cham3}. Besides,
it has been confirmed that the original setting of the mimetic
theory predicts that gravitational wave (GW) propagates at the
speed of light, ensuring agreement with the results of the event
GW170817 and its optical counterpart \cite{Sunny1,Sunny2}. It has
also been shown that this theory can explain the flat rotation
curves of spiral galaxies \cite{MimMOND}. Mimetic theory of
gravity has arisen a lot of enthusiasm in the past few years both
from the cosmological viewpoint \cite{MimCos,
Baf,Dutta,Sep,Mat,Leb,MimCos2,Gorj1,Gorj2,Gorj3,Gorj4,Fir,Cham2,Russ,Ces,Vic,Fir2}
as well as black holes physics \cite{Der,Myr1,Myr2,Alex,
Chen,Nash3,jibril,Bra,Yunlong1,Yunlong2,Nash1,Cham4,Gorji2,Shey1,Shey2,NashNoj,ACham,YZ,Bakh,Nash4}.
The studies on mimetic gravity were also generalized to $f(R)$
gravity
\cite{Odin0,Odin1,Oik1,Oik2,Oik3,MyrzSeb,OdinP,OdinOik,Odin3,Odinfr1,Odinfr2,Bhat,Adam,Jing}
and Gauss-Bonnet gravity \cite{OdinGB,Oik,Zh1,Zh2,Paul}. In particular, a unified description of early inflation and late-time acceleration in the context of
 mimetic $F(R)$ gravity was established in \cite{Odin2}. It has been confirmed that in the background of mimetic $F(R)$ gravity, the inflationary era
can be realized \cite{Odin2}. This is in contrast to the usual $F(R)$ gravity where it suffers to explain the inflationary era.

On the other side, it is well-known that the topology of the event
horizon of an asymptotically flat stationary black hole in four
dimensions is uniquely determined to be the two-sphere $S^2$
\cite{Haw1,Haw2}. Indeed, Hawking's theorem requires the
integrated Ricci scalar curvature with respect to the induced
metric on the event horizon to be positive \cite{Haw1}. Of course,
when the spacetime is not asymptotically flat, the spherical
topology of the black holes horizon is not necessary and one can
have stationary black holes with nontrivial topologies. It was
argued that for asymptotically AdS spacetime, in the
four-dimensional Einstein-Maxwell theory, there exist black hole
solutions whose event horizons may have zero or negative constant
curvature and their topologies are no longer the two-sphere $S^2$
(see e.g. \cite{Lemos,Cai22,Bril1,Cai3,Cai4,Shey3,Shey4,Shey5} and references
therein).

Despite a lot of efforts in exploring the mimetic gravity, one
important question remains to be answered. Is it possible to
reproduce the flat galactic rotation curves, in a static
spacetime, in mimetic gravity?  Our aim in this work is to
construct topological black hole solutions in the context of
mimetic gravity. For completeness, we study the case of uncharged
and charged black holes in the absence and presence of a constant
potential for the mimetic field. We find that in order to fully
satisfy the field equations, the mimetic potential should be
regarded as a negative constant which can admit an asymptotically
anti-de Sitter (AdS) spacetime. We also calculate the orbital
velocity of a test particle in mimetic spacetime and show that it
can serve as an alternative explanation of the flat rotation
curves of spiral galaxies. Our work differs from \cite{MimMOND} in
that we consider the original version of mimetic theory introduced
in \cite{Mim1} to explain the flat galactic rotation curves, while
the author of \cite{MimMOND} reproduces the MOND-like theory
within the framework of mimetic gravity, by adding a non-minimal
coupling between the mimetic field and the matter hydrodynamic
flux. Our work also differs from \cite{Myr2} where by implementing
linear and quadratic corrections to the potential of the mimetic
field, they modified the effective Newtonian gravitational
potential felt by a test particle. Here, we disclose that in order
to explain the flat galactic rotation curves in mimetic gravity,
one does not need to add neither a non-minimal coupling term to
the action nor taking into account a potential for the mimetic
field. We will also analyze our solutions by exploring the orbits
of test particles and light, which is a powerful method to study
the structure and properties of a spacetime. With the help of
geodesics observable quantities like  the shadow of a black hole
can be calculated and compared to observations. This allows to
test different models and theories of gravity. We will derive the
equations of motion using the Hamilton-Jacobi approach and study
the behaviour of the geodesics with effective potential
techniques.

This paper is structured as follows. In the next section, by
varying the action of mimetic gravity, we derive the basic field
equations. In section \ref{uncharge}, we obtain topological black
hole solutions in mimetic gravity and study the physical
properties, asymptotic behaviour and casual structure of the
solutions. We also obtain the orbital speed of test particle in
mimetic spacetime and show that how it can alleviate the problem
of flat rotation curves of spiral galaxies. In section
\ref{charge}, we extend our study to charged topological black
holes in the context of mimetic gravity. In section
\ref{geodesic}, we explore the geodesic motion of massless and
massive test particles in these spacetimes. We finish with closing
remarks in the last section.
\section{Basic Field Equations}\label{Field}
We start with the following action
\begin{eqnarray}\label{Act}
S=\int{ d^4x\sqrt{-{g}}\left[R+{\lambda}( g^{\mu \nu}\partial
_{\mu} \phi \partial _{\nu}\phi-\epsilon)- V(\phi)+
\mathcal{L}_{m}\right]},
\end{eqnarray}
where $R=R(g_{\mu \nu})$ is the Ricci scalar, $V(\phi)$  is the
potential for the scalar field, $\lambda$ is the Lagrange
multiplier, and $\mathcal{L}_{m}=-F_{\mu \nu } F^{\mu \nu }$ is
the Lagrangian of the Maxwell field. Here $F_{\mu \nu }=\partial
_{\mu }A_{\nu }-\partial _{\nu }A_{\mu }$ is the electromagnetic
field tensor and $A_{\mu }$ is the gauge potential, and
$\epsilon=\pm1$ depends on the spacelike or timelike nature of
$\partial _{\mu} \phi$. Through this paper we take the signature
$(-,+,+,+)$, thus $\epsilon=-1$ corresponds to timelike $\partial
_{\mu} \phi$ and $\epsilon=1$ indicates spacelike $\partial _{\mu}
\phi$.  We further assume $8\pi G_N=1$, unless we mention
explicitly. Varying the above action with respect to the metric
$g_{\mu \nu}$ and the scalar field $\phi$, leads to the following
equations of motion
\begin{eqnarray}\label{FE1}
{G}_{\mu\nu}&=& \lambda \partial _{\mu} \phi \partial _{\nu}
\phi-\frac{1}{2}g_{\mu \nu }V(\phi)+T_{\mu \nu }=T^{\phi}_{\mu \nu
}+T_{\mu \nu }
\end{eqnarray}
\begin{equation}\label{FE2}
\nabla ^{\mu}(\lambda \partial _{\mu} \phi )=-\frac{1}{2}\frac{d V
(\phi)}{d \phi},
\end{equation}
where the energy momentum tensor of the scalar and electromagnetic
field are defined,
\begin{eqnarray}\label{Tphi}
T^{\phi}_{\mu \nu }&=&\lambda \partial _{\mu} \phi \partial _{\nu}
\phi-\frac{1}{2}g_{\mu \nu }V(\phi),\\
 T_{\mu \nu }&=&2 F_{\mu
\gamma } F_{ \nu }^{\ \gamma}-\frac{1}{2} g_{\mu \nu } F_{\alpha
\beta } F^{\alpha \beta },\label{Tem}
\end{eqnarray}
while the equation of motion for the gauge field can be obtained
as
\begin{equation}
\partial _{\mu }\left( \sqrt{-g} F^{\mu \nu }\right)
=0. \label{FE3}
\end{equation}%
Variation of the action (\ref{Act}) with respect to the Lagrange
multiplier $\lambda$ leads
\begin{equation}\label{cond}
g^{\mu \nu}\partial _{\mu} \phi \partial _{\nu} \phi=\epsilon.
\end{equation}
This implies that the scalar field is always constrained by Eq.
(\ref{cond}) and thus is not dynamical by itself, but induces
mimetic dark matter in Einstein theory making the longitudinal
degree of freedom of the gravitational field dynamical
\cite{MimCos1}. The mimetic gravity proposed in \cite{Mim1}
corresponds to $\epsilon=-1$ which means that $\partial _{\mu}
\phi$ should be timelike (their signature differs from ours). We
emphasize that introducing $\phi$ does not add a new dynamical
scalar field. This is the main difference between mimetic gravity
and standard scalar-tensor theories of gravity such as Brans-Dicke
or dilaton gravity where the scalar field adds a new degree of
freedom. In contrast, in mimetic gravity, in addition to two
transverse degrees of freedom describing gravitons, the
gravitational field acquires an extra longitudinal degree of
freedom induced by the mimetic field \cite{Mim1}. Thus, mimetic
gravity can be viewed as a modification of General Relativity in
the longitudinal sector.

Taking the trace of Eq. (\ref{FE1}) and using the constraint
equation (\ref{cond}), we find
\begin{eqnarray}\label{lambda}
\lambda=\epsilon(G-T+2 V),
\end{eqnarray}
where $G=G^{\mu}_{\mu}$ and $T=T^{\mu}_{\mu}$ are, respectively,
the trace of the Einstein tensor and energy momentum tensor.
Inserting $\lambda$ in the field Eqs. (\ref {FE1}) and (\ref
{FE2}), we arrive at
\begin{eqnarray}\label{FEE1}
{G}_{\mu\nu}&=& \epsilon(G-T+2V) \partial _{\mu} \phi \partial
_{\nu} \phi-\frac{1}{2}g_{\mu \nu }V(\phi)+T_{\mu \nu }
\end{eqnarray}
\begin{equation}\label{FEE2}
\nabla ^{\mu}[(G-T+2V)\partial _{\mu}
\phi]=-\frac{\epsilon}{2}\frac{d V (\phi)}{d \phi}.
\end{equation}
Our aim here is to find the topological static black hole
solutions of the above field equations. We assume the line
elements of the metric as
\begin{equation}\label{metric}
ds^2=- g^2(r)f(r)dt^2 +{dr^2\over f(r)}+ r^2 d\Omega_{k}^2 ,
\end{equation}
where $f(r)$ and $g(r)$ are unknown functions of $r$ which should
be determined, and $d\Omega_{k}^2$ is the line element of a
two-dimensional hypersurface $\Sigma$ with constant curvature,
\begin{equation}\label{met}
d\Omega_k^2=\left\{
  \begin{array}{ll}
    $$d\theta^2+\sin^2\theta d\varphi^2$$,\quad \quad\!\!{\rm for}\quad $$k=1$$, &  \\
    $$d\theta^2+d\varphi^2$$,\quad\quad\quad\quad {\rm for}  \quad $$k=0$$,&  \\
    $$d\theta^2+\sinh^2\theta d\varphi^2$$, \quad {\rm for}\quad $$k=-1$$.&
  \end{array}
\right.
\end{equation}
For $k = 1$, the topology of the event horizon is the two-sphere
$S^2$, and the spacetime has the topology $R^2 \times S^2$. For $k
= 0$, the topology of the event horizon is that of a torus and the
spacetime has the topology $R^2 \times T^2$. For $k = -1$, the
surface $\Sigma$ is a $2$-dimensional hypersurface $H^2$ with
constant negative curvature. In this case the topology of
spacetime is $R^2 \times H^2$.

Using the metric (\ref{metric}), the constraint equation
(\ref{cond}) yields
\begin{equation}\label{phi}
f(r) \phi'^{2}=\epsilon,   \   \rightarrow   \   \phi(r)=\pm
\int{\frac{dr}{\sqrt{\epsilon f(r)}}}+\rm const.,
\end{equation}
which again confirms that the scalar field $\phi$ is not
dynamical. For timelike $\partial _{\mu} \phi$ ($\epsilon=-1$),
the scalar field is imaginary. However, since $\phi$ is not a
dynamical field, it does not make any trouble and we have still a
real spacetime. Suppose the gauge potential is in the form
$A_{\mu}=h(r) \delta_\mu^{0}$, and using the metric
(\ref{metric}), the only non-vanishing component of Eq.
(\ref{FE3}) is given by
\begin{equation}\label{hr}
-rE(r)g'(r)+2E(r)g(r)+rE'(r)g(r)=0,
\end{equation}
where $F_{tr}=E(r)=h'(r)$ is the electric field and the prime
denotes the derivative with respect to $r$. Solving the above equation
for $E(r)$, we find
\begin{equation}\label{Er}
E(r)=\frac{q}{r^2}g(r),
\end{equation}
where $q$, which is a constant of integration, is related to the
electric charge of the black hole. Inserting metric (\ref{metric})
and the electric field (\ref{Er}) into the field equations
(\ref{FEE1}), regardless the sign of $\epsilon$, we obtain the
following equations for the components of the Einstein equations,
\begin{eqnarray}\label{tt}
&&r^3 f^{\prime}-k{r}^{2}+f {r}^{2}+{r}^{4}{V \left( \phi
\right)}/2+{q}^{2}=0,\\
&&3\, {r}^{3} g f^{\prime}+2\, {r}^{3}f g^{\prime}-k {r}^{2} g
+{r}^{2} g
  f +3\, {r}^{4} f^{\prime}g^{\prime} +2\, {r}^{4} f  g'' +{r}^{4} g  f''
  +3{r}^{4} g
 V(\phi)/2-{q}^{2}g =0,\label{rr}\\
&& 2\, {r}^{3} (f  g)' +3 {r}^{4}f' g' +2 {r}^{4} f
 g'' +{r}^{4} g f'' +{r}^{4} \,g
 V \left( \phi \right)-2\,{q}^{2}g =0.
\label{pp}
\end{eqnarray}
In the remaining part of this paper, we are going to solve the
above field equations and obtain the unknown functions $f(r)$ and
$g(r)$. Clearly, our solutions should also satisfy Eq.
(\ref{FEE2}) for the scalar field. Through this paper, we take two
values for the potential of the mimetic field, namely $V(\phi)=0$
and $V(\phi)=-V_0=-2\Lambda$ with $\Lambda>0$. We also find out
that our solutions do not exist for the case of a positive
constant potential. Besides, when the mimetic potential is a
function of $r$, namely for $V(r)=V[\phi(r)]$, it is not easy to
find an exact analytical solution for the full field equations.
Although in \cite{Myr2}, the authors considered several type of
variable potentials and presented some \textit{approximate}
solutions, but their solutions are not the exact solutions of the
full field equations and could be only valid for some range of the
distance $r$. Thus, we leave the solutions of the topological
mimetic black holes for a variable potential for future studies.
\section{Uncharged black holes} \label{uncharge}
In this section, we consider the case where our black hole
solutions have no charge by setting $q=0$ in the field equations
(\ref{tt})-(\ref{rr}).
\subsection{Solutions with $V(\phi)=0$}
At first we investigate the case where $V(\phi)=0$. In this case
the field equations reduce to
\begin{eqnarray}\label{ttun}
&&rf^{\prime}+f-k =0,\\
&&3\, {r} g f^{\prime}+2\, {r}f g^{\prime}-k g + g
  f +3\, {r}^{2} f^{\prime}g^{\prime} +2\, {r}^{2} f  g'' +{r}^{2} g  f''
   =0,\label{rrun}\\
&& 2\, (f  g)' +3 {r}f' g'  +2 {r} f
 g'' +{r} g f'' =0.
\label{ppun}
\end{eqnarray}
The solution to Eq. (\ref{ttun}) is given by
\begin{eqnarray}\label{frun}
f(r)=k-\frac{m}{r}.
\end{eqnarray}
Inserting this solution into Eq. (\ref{rrun}), we  find the
following solution for the metric function
\begin{eqnarray}\label{grun}
g(r)&=& 1+b_0 \left[-2 \left(1-\frac {km}{r}\right)^{-1/2}+\ln
\left( \frac{r}{r_0}+\frac{r}{r_0}\sqrt{ 1-{\frac {km}{r}}}-\frac
{km}{2r_0} \right)\right], \quad\!\!{\rm for}\quad k=\pm 1,\\
g(r)&=&1+b_1 r^{3/2}  \quad\!\!{\rm for}\quad k=0, \label{grun2}
\end{eqnarray}
where $b_0$, $b_1$ and $r_0$ are integration constants. One can
easily check that solutions (\ref{phi}), (\ref{frun}) and
(\ref{grun}) also satisfy the remaining field equations
(\ref{FEE2}) and (\ref{ppun}). The black hole horizon can be
obtained from equation $\textbf{\textbf{g}}^{rr}=f(r)=0$, which
has a solution only for the case $k=+1$. For $k=0,-1$ the
solutions do not represent a black hole and we encounter a naked
singularity. In order to have a better insight to the nature of
the solution, let us calculate the $(tt)$ component of the metric
function. From the line element (\ref{metric}), we have
$\textbf{g}_{tt}=-B(r)=-f(r)g^2(r)$, where
\begin{eqnarray}\label{g00un}
&&B(r)= \left(k-\frac{m}{r}\right) \Bigg{\{} 1+b_0 \left[-2
\left(1-\frac {km}{r}\right)^{-1/2}+\ln \left(
\frac{r}{r_0}+\frac{r}{r_0}\sqrt{ 1-{\frac {km}{r}}}-\frac
{km}{2r_0} \right)\right] \Bigg{\}}^2 ,
{\rm for}\quad k=\pm 1, \nonumber\\
&&B(r)= -\frac{m}{r}\left(1+b_1 r^{3/2}\right)^2 \quad\!\!{\rm
for}\quad k=0 \, .
\end{eqnarray}
Clearly for $k=0,-1$ we have always $\textbf{g}_{tt}>0$, which
means the signature of the metric is changed. Again, we confirm
that, in mimetic gravity similar to Einstein gravity, we have no
topological black holes in the case of zero potential and the horizon
topology must be a $2$-sphere.

In order to check the curvature singularity of the spacetime, we
examine the Ricci and Kretschmann invariants for the obtained
solutions. It is easy to show that these invariants diverge at
$r=0$, they are finite at $r\neq 0$ and go to zero for
$r\rightarrow \infty$. Thus there is a curvature singularity
located at $r=0$.

It is also interesting to study the behaviour of the metric
function at far distances. Of course, we are aware that in the
weak gravitational field regime, $B(r)\approx1+2\Phi (r)/c^2$,
where $\Phi(r)$ is the Newtonian gravitational potential and $c$
is the speed of light. One also knows that the speed of an
orbiting test object in this spacetime is given by
\begin{eqnarray}\label{v2}
v^2(r)= r \frac{d \Phi(r)}{dr}=\frac{1}{2} r c^2 \frac{d
B(r)}{dr}.
\end{eqnarray}
In order to have a better insight on the above behaviour of the
circular speed, let us expand $B(r)$ given in (\ref{g00un}) for
large values of $r$, namely at far distance compared to the
horizon radius, $r\gg m$. It is a matter of calculation to show
that($k=1$)
\begin{eqnarray}\label{Brex}
B(r)&\approx & c_1-\frac{m}{r}+c_3 \ln r+c_4 (\ln r)^2+ c_5
\frac{m\ln r}{r}+ c_6 \frac{m(\ln r)^2}{r}+
O\left(\frac{1}{r^2}\right).
\end{eqnarray}
where $c_i=c_i(b_0,r_0)$ are constants such that in the absence of
a mimetic field ($b_0=0$) we have $c_i=0$ for $i\geq 3$. The
constant $c_1$ can be absorbed in a redefinition of the time
coordinate $t$ and hence can be set equal to one. Note that the
higher order terms are all in the form $1/r^n$ with $(n\geq2)$,
and there is no contribution from the logarithmic term in the
higher order terms. As we shall see, the logarithmic terms play a
crucial role in the behaviour of the orbital speed of a test
particle. Without the logarithmic terms, the orbital speed would
be a decreasing function of $r$ and the flat galactic rotation
curves are remained upset.

\begin{figure}[h]
  \centering
  \subfigure[$c_4<1$]{
    \includegraphics[width=0.47\linewidth]{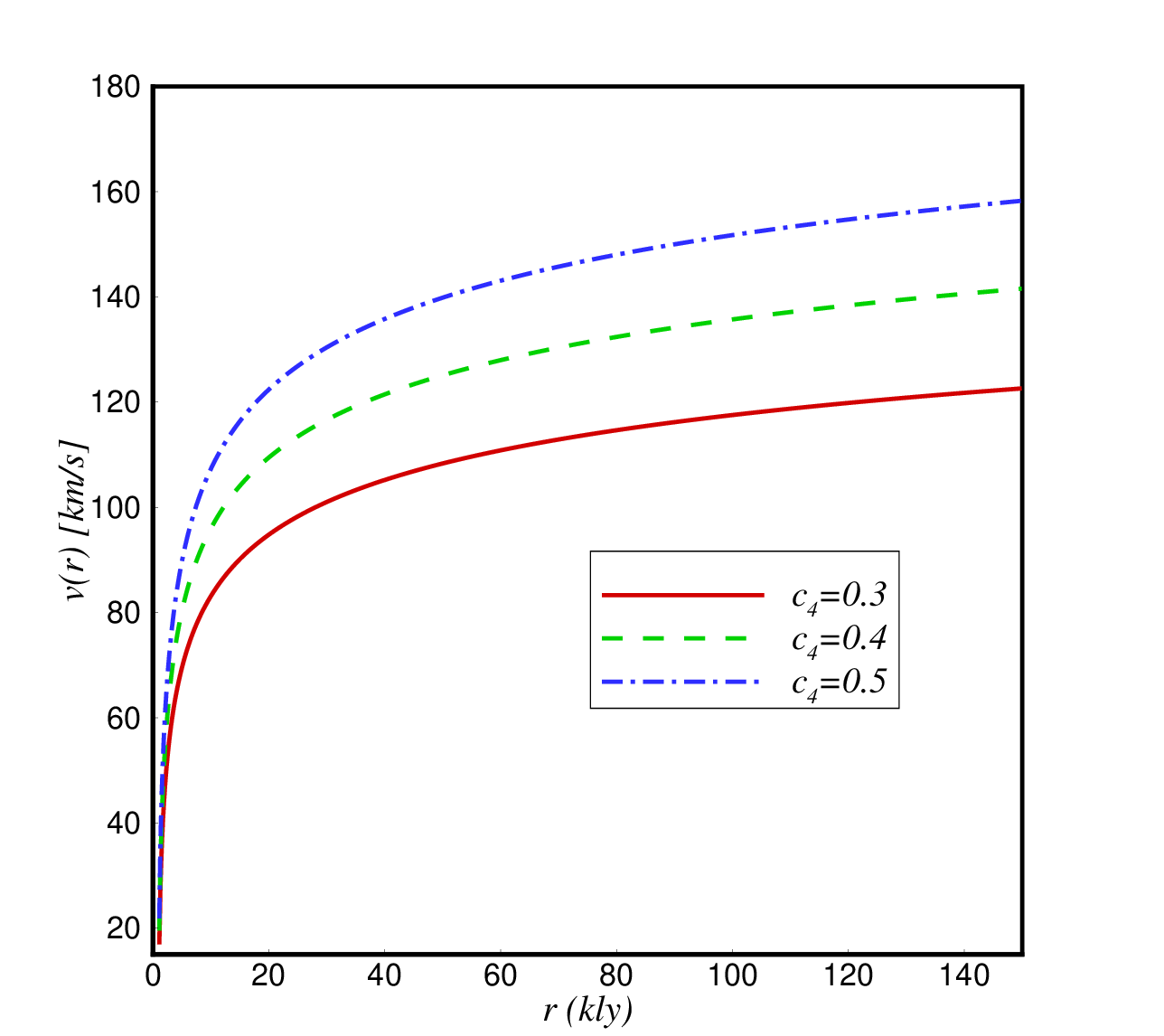}
  }
  \subfigure[$c_4>1$]{
    \includegraphics[width=0.47\linewidth]{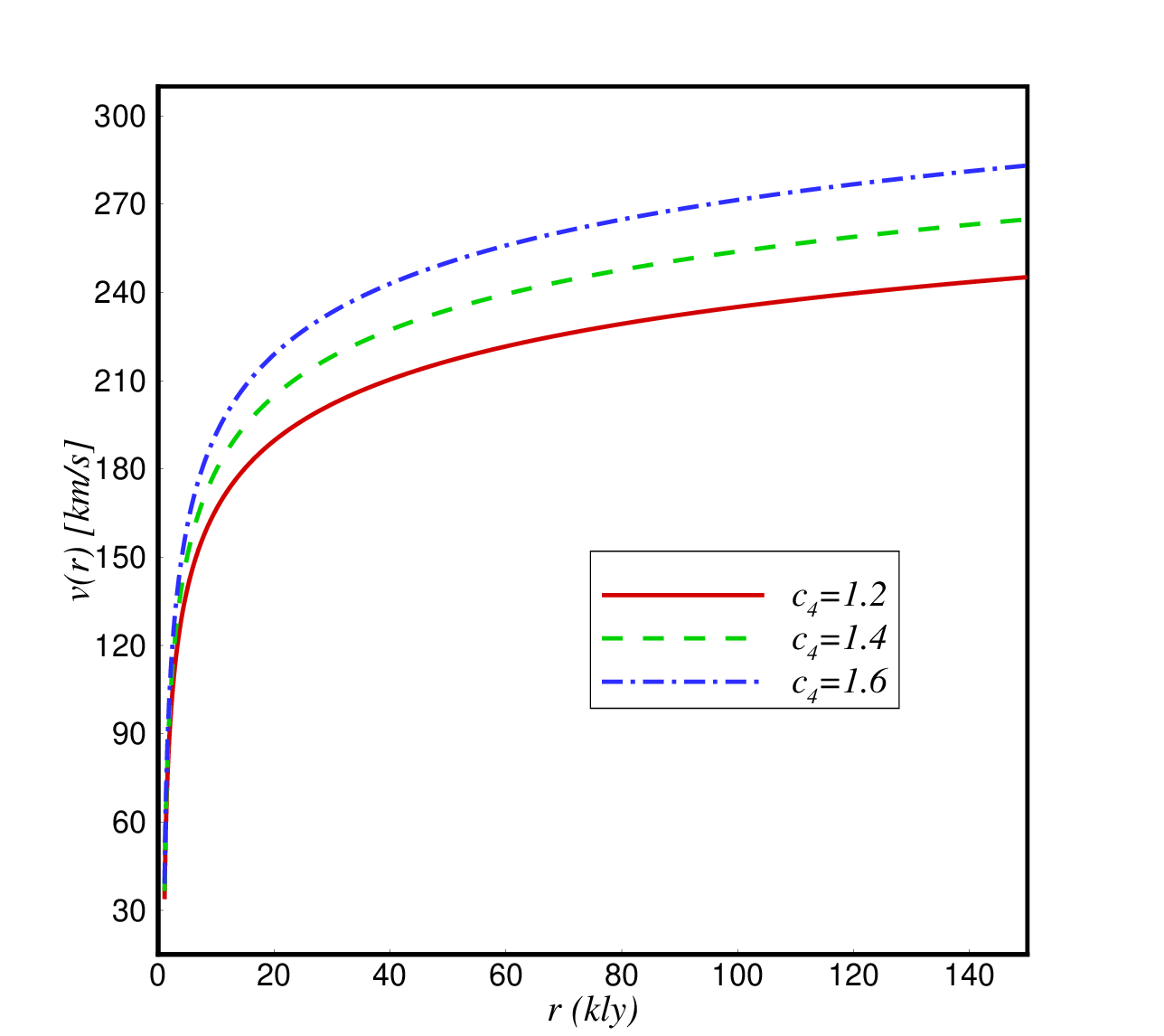}
  }

  \caption{The circular speed
of an orbiting test particle around a typical spiral galaxy with
mass $M=10^{12}M_{\odot}$  in the mimetic spacetime in terms of
distance $r$. Here, we have taken $c_5=c_6=0$ and kept fixed
$c_3=2.8 \times 10^{-12} \sqrt{M/M_{\odot}}=2.8 \times 10^{-6}$
\cite{Sobuti}.}
  \label{Fig1ab}
\end{figure}

\begin{figure}[h]
  \centering
  \subfigure[different $c_4$]{
    \includegraphics[width=0.47\linewidth]{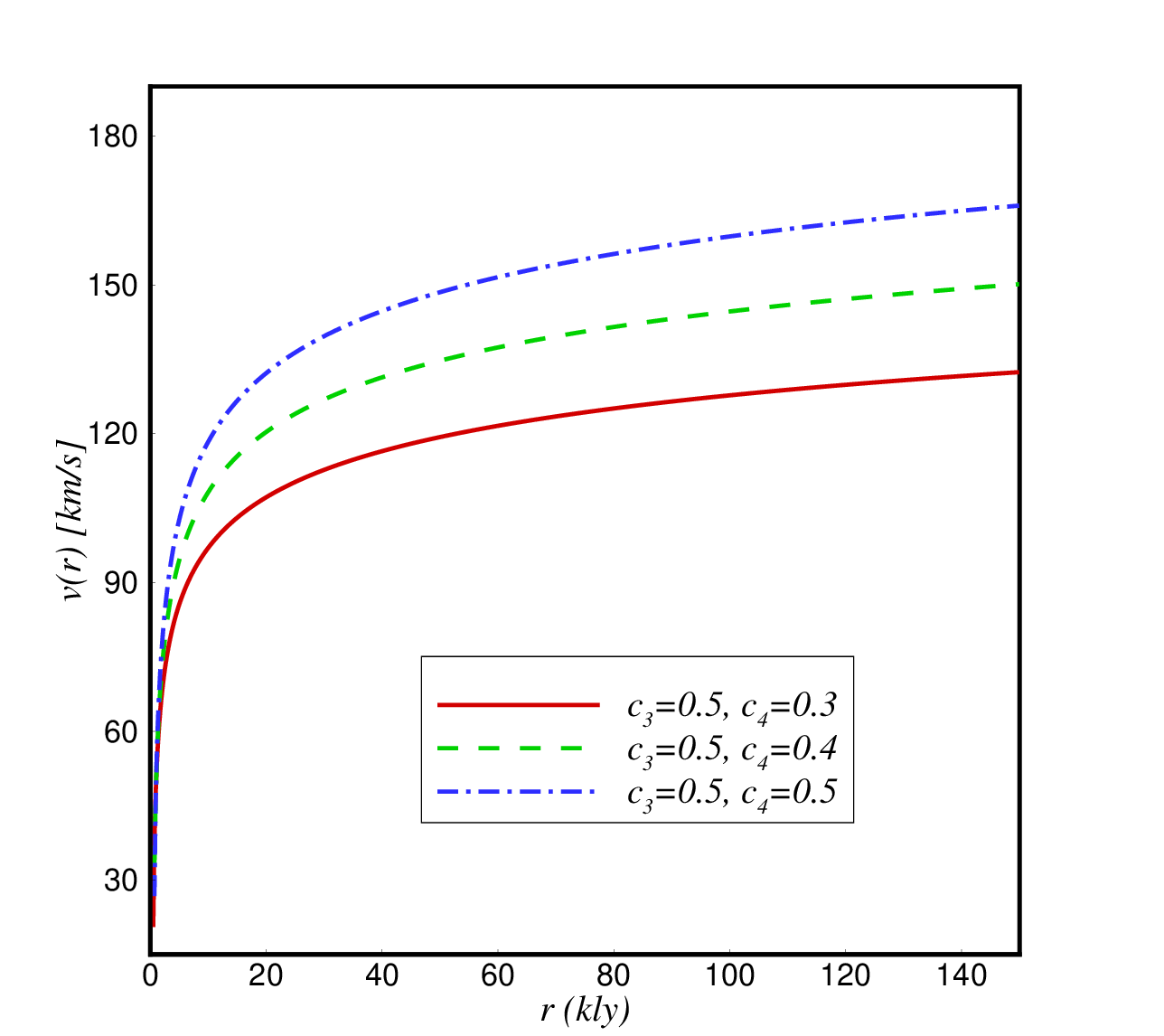}
  }
  \subfigure[different $c_3$]{
    \includegraphics[width=0.47\linewidth]{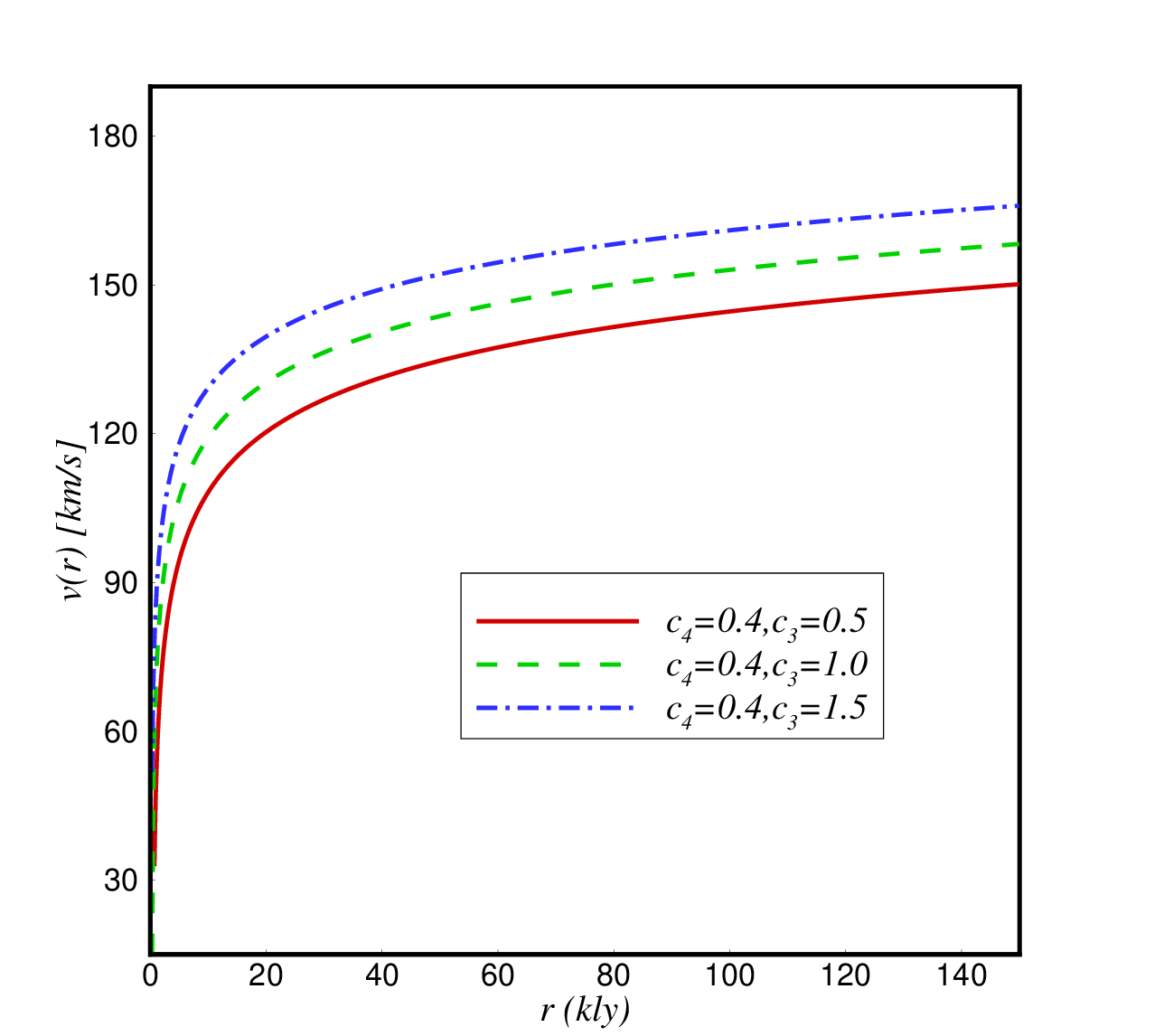}
  }
  \caption{The circular speed
of an orbiting test particle around a typical spiral galaxy with
mass $M=10^{12}M_{\odot}$  in mimetic spacetime in terms of
distance $r$. Here, we have taken $c_5=c_6=2$.}
  \label{Fig1cd}
\end{figure}

Substituting $B(r)$ from Eq. (\ref{Brex}) into Eq. (\ref{v2}) one
can obtain the functional form of the circular speed $v(r)$ in
terms of $r$ which depends on the parameters $c_i$ and $m$.
However, it is more instructive to plot $v(r)$ for different
values of the parameters. The behaviour of the orbital speed of a
test particle in this spacetime, can be applied for understanding
the flat rotation curves of spiral galaxies. This can be achieved
by assuming that the underlying theory which describe the
spacetime around a galaxy is the mimetic gravity. It is well-known
that the circular velocity of spiral galaxies at far distances, at
the galaxy outskirt, tends to a constant value. Let us note that
in the limiting case where $b_0=0$, our spacetime reduces to the
Schwarzschild one, therefore we realize that $m=2G_{N}M/c^2$.
Besides, for $c_5=c_6=0$, our solution (\ref{Brex}) restores the
one proposed in \cite{Sobuti}, which implies that, in this case we
can define $c_3=\lambda_0=2.8\times 10^{-12}\sqrt{M/M_{\odot}}$
(see Eq. (\ref{BrSob})). In order to be more realistic, we
consider a typical spiral galaxy with mass $M=10^{12} M_{\odot}$,
and thus $m=2\times 10^{12}G_{N} M_{\odot}/c^2$, where
$G_{N}=6.674\times 10^{-11} m^3 kg^{-1}s^{-2}$ is the Newtonian
gravitational constant and $M_{\odot}\approx 10^{30} kg$ is the
mass of the Sun. Substituting $m$ in (\ref{Brex}), with
appropriate choice of the other parameters, we can plot the
orbital speed in terms of the distance $r$ from the galaxy center.
We summarize our results in Figs. \ref{Fig1ab} and \ref{Fig1cd}
where $r$ is given in unit of $``kly"$. These figures are
compatible with astrophysical data \cite{NP,Man,Bri}, although we
are not going to give the details of data fitting of our model
with observations. Indeed, here we present the ideas, and show how
mimetic gravity can provide a theoretical basis for explaining the
flat rotation curves of spiral galaxies. The detailed data fitting
of the parameters space with observations is left for future
studies.

From these figures we observe that the orbital speed crucially
depends on the parameters $c_3$ and $c_4$ but is not sensitive
with respect to the parameters $c_5$ and $c_6$. From Fig.
\ref{Fig1ab}, we see that even for $c_5=c_6=0$ we still have a
desired flat galactic rotation curve. We observe that the orbital
speed increases for small distances and tends to a constant value
at large distances, compatible with astrophysical data
\cite{NP,Man,Bri}. Besides, at any distance, the orbital speed
$v(r)$ increases with increasing the parameter $c_4$ which
incorporates the effects of mimetic gravity. The reason for this
behaviour originates from the fact that with increasing $c_4$, the
logarithmic term which comes from mimetic field in the solution
(\ref{Brex}) grows up. In Fig. \ref{Fig1cd}, however, we keep
$c_5$ and $c_6$ fixed and allow the parameters $c_3$ and $c_4$ to
vary. Obviously, with increasing either $c_3$ or $c_4$, the
orbital speed increases as well. Of course, when $b_0=0$
($c_i=0$), namely for the Schwarzschild spacetime, the orbital
speed restores $v(r)=c\sqrt{m/2r}=\sqrt{G_{N}M/r}$, which is a
decreasing function of $r$. Thus, the impact of the mimetic
gravity dramatically changes the behaviour of the orbital speed of
a test particle in this spectime.

{A question then may arise: what is the origin of such behaviour
of the metric function? In other words, why the spectime metric
can naturally explain the flat rotation curves in mimetic gravity?
In order to address this question, let us recall the original work
\cite{Mim1} where the idea of \textit{mimetic dark matter} was
initiated. In \cite{Mim1}, the authors argued that, on the flat
FRW Universe where the mimetic field is a function of time
$\phi=t$, the energy density of the mimetic field can be written
in the form $\rho_{mim}=G-T=C(x^i)/a^3$ where $a$ is the scale
factor of the Universe and, thus mimicking the contribution of
pressureless dust. The constant $C(x^i)$ determines the value of
the mimetic dark matter. Even, in the absence of matter where
$T=0$, this mimetic energy density is equal to $\rho_{mim}=-R$,
which, in general, has a non-vanishing value. This implies that,
in mimetic gravity, dark matter appears naturally as a purely
geometrical effect.} This mimetic energy density comes from the
extra longitudinal degree of freedom of the gravitational field
equation induced by the mimetic field. It is worth noticing that,
for consistency, in this picture, one should define the
four-velocity of this mimetic dust as $u^{\mu}=g^{\mu
\nu}\partial_{\nu} \phi$ which satisfies the normalization
condition $u^{\mu} u_{\mu}=\epsilon$. This means the mimetic filed
plays the role of the velocity potential. Clearly, the
normalization condition for the four-velocity is equivalent to the
condition (\ref{cond}) for the mimetic field $\phi$. Now we back
to the above question. In the context of static spacetime where
$\phi=\phi(r)$, the field equations admit a solution which can
explain the flat galactic rotation curves, but how? Indeed the
responsible term for such behaviour is the second term in the
metric function (\ref{grun}), namely $g(r)-1$. This term in
solution (\ref{grun}) contributes from the mimetic term
$(G-T+2V)\partial_{\mu}\phi
\partial_{\nu} \phi$ in the gravitational field Eq. (\ref{FEE1}),
which is indeed the energy-momentum tensor of the mimetic field.
Note that in our case, $V=0=T$ and this term is reduced to
$-R\partial_{\mu}\phi
\partial_{\nu} \phi$.
Without this mimetic term, the field equations admit a unique
solution $g(r)=1$ and necessary implies $b_0=0$, which is just the
solution of General Relativity.

In order to clarify how the second term in $g(r)$ reflects the
effect of mimetic field, let us note that the functional form of
$g(r)$ in (\ref{grun}), before integrating, is given by
\begin{eqnarray}\label{grunb}
g(r)&=& 1+b_0 \int{\frac{\sqrt{r} \ dr}{(r-m)^{3/2}}}=1+b_0
\int{\frac{dr}{r
[f(r)]^{3/2}}}=1+b_0\int{r^{-1}[\epsilon\phi'(r)]^{3}dr}.
\end{eqnarray}
Obviously, in the absence of mimetic field ($\phi=0$), the second
term vanishes and our solution reduces to that of General
Relativity. In conclusion, the origin of the responsible term,
which is capable to reproduce the flat galactic rotation curves in
a static spacetime, is the one which mimics \textit{dark matter}
in the cosmological setup \cite{Mim1}. This is the reason why the
velocity of a test particle in mimetic gravity, in a static
spacetime, can be naturally explained and the inferred flat
galactic rotation curves are reproduced without invoking particle
dark matter. More precisely, the proposed model of mimetic theory
of gravity in \cite{Mim1} is defined in such a way that naturally
mimics dark matter as a geometrical effect. Therefore, it is not
strange that we could reproduce the flat galactic rotation curves
in the background of static spacetime in mimetic gravity, without
any modification in the action \cite{MimMOND} or taking into
account a variable potential \cite{Myr2}.

Our work may also provide a theoretical origin for the proposed
modification of the Schwarzschild spacetime in Ref. \cite{Sobuti}
for the explanation of the flat galactic rotation curves. Let us
recall that, in order to explain the flat rotation curves of
spiral galaxies, the author of \cite{Sobuti} proposed that the
flat rotation curves of spiral galaxies can be explained by
logarithmic gravitational potentials. He assumed the $(00)$
component of the metric around a galaxy, at far distance, in
addition to the Schwarzschild term, should have a logarithmic
term,
\begin{eqnarray}\label{BrSob}
B(r)=1-\frac{r_s}{r}+ \lambda_0 \ln r.
\end{eqnarray}
where $r_s=2G_{N}M/c^2$ is the Schwarzschild radius of the galaxy,
and $\lambda_0 \approx 2.8\times 10^{-12} \sqrt{M/M_{\odot}}$
\cite{Sobuti}. In order to justify the appearance of the
logarithmic term in the metric function around a galaxy, the
author \cite{Sobuti} modified the underlying theory of gravity by
adding a new energy momentum tensor to Einstein's field
equations. He proposed that a spiral galaxy has a dark perfect
fluid companion which provides such modification in the metric.
Here, we observe that the mimetic gravity can serve a logarithmic
term in the metric function without needing to add a dark companion
to the galaxy matter. Indeed, as argued in \cite{Mim1}, the extra
longitudinal degree of freedom of the gravitation field, in
mimetic gravity, is responsible for the appearance of such a
logarithmic term in the metric function which could have naturally
appeared at far distances. Thus, in our work the dark matter can
be understood as a geometrical impact. Finally, we emphasize that
in Eq. (\ref{Brex}), we present the next correction terms to the
Schwarzschild metric which comes from the longitudinal degree of
freedom of the gravitational field in mimetic gravity.
\subsection{Solutions with $V(\phi)=-2\Lambda$}
Next, we consider the case in the presence of a constant potential
for the scalar field. Indeed, the consistent solutions, which
fully satisfy the field equations, only exist for the case
$\Lambda>0$. In this case the field Eqs. (\ref{tt})-(\ref{pp})
with $q=0$, admit the following solutions
\begin{eqnarray}\label{fr2un}
f(r)&=&k-\frac{m}{r}+\frac{\Lambda r^2}{3}, \quad\ \ \ \quad\ \quad\  \quad\ \quad\ \quad\!\!{\rm for}\quad k=0,\pm 1, \\
g(r)&=&1+b_2 \frac{r^{3/2}}{\sqrt {\Lambda r^3-3m}} \ \ \quad\
\quad\ \quad\ \quad\ \quad\!\!{\rm
for}\quad k=0,\label{gr02un} \\
g(r)&=& 1+ c_0\int{\frac{\sqrt{r} \  dr}{(\Lambda r^3
-3m+3kr)^{3/2}}}, \quad\ {\rm for}\quad k=\pm 1 \, .
 \label{gr2un}
\end{eqnarray}
The horizon can be obtained from $g^{rr}=f(r)=0$. For $k=0$, black
hole has a single horizon located at $r_h=(3m/\Lambda)^{1/3}$, and
from (\ref{gr02un}) we observe that $g(r)\rightarrow \infty$ at
$r=r_h$. However, this is just a coordinate singularity and all
curvature invariants are finite at $r=r_h$. For $k=\pm1$, however,
the horizon is the real positive root of $\Lambda r^3 -3m+3kr=0$.
In these cases, we have again one horizon located at $r_{+}$ where
its radius depends on the parameters $m$ and $\Lambda$. We have
also a coordinate singularity at $r_{+}$, as one can see from
solution (\ref{gr2un}).

Again, it is a matter of calculations to show that these solutions
together with (\ref{phi}) fully satisfy the field equation for the
scalar field given in (\ref{FEE2}). Since the integral in Eq.
(\ref{gr2un}), cannot be done analytically, we expand the
integrand for large values of $r$. We find
 \begin{eqnarray}\label{grunexp}
g(r)\approx1+ \frac{c_0}{\Lambda^{5/2} r^5} \Bigg{\{}
\frac{9k}{10}-\frac{\Lambda r^2}{3}-\frac{3m}{4r}+
O\left(\frac{1}{r^{2}}\right)\Bigg{\}}, \quad\ {\rm for}\quad
k=\pm 1 \label{gr3exp} \, .
\end{eqnarray}
Similarly, the function $B(r)=f(r)g^2(r)=-\textbf{g}_{tt}$ at
large distance is obtained
\begin{eqnarray}\label{gttexp1}
B(r)&=& \frac{\left(\sqrt{\Lambda r^3 -3m}+b_2
r^{3/2}\right)^2}{3r} \nonumber \\ &\approx& \frac{r^2}{3}\left(
\Lambda+b_2^2+2b_2 \sqrt{\Lambda}\right)-\frac{m}{r}
\left(1+\frac{b_2}{\sqrt{\Lambda}}\right)+O\left(\frac{1}{r^4}\right)
\ \ {\rm for}\ k=0,\\
 B(r)&\approx
&k-\frac{m}{r}+\frac{\Lambda r^2}{3}-\frac{2c_0}{9\sqrt{\Lambda}}
\frac{1}{r}+ O\left(\frac{1}{r^{3}}\right) \quad\ {\rm for}\quad
k=\pm 1. \label{gttexp2}
\end{eqnarray}
We observe that for a flat horizon $(k=0)$, $B(r)$ has a minimum
value $B(r)\mid_{\rm min}=b_2^2 r_{h}^2/3=b_2^2
(3m/\Lambda)^{2/3}/3$ at the horizon. Besides, from solution
(\ref{gttexp1}) we see that, for $k=0$, the asymptotic behaviour
is \textit{approximately} AdS, unless in the absence of the
mimetic field $(b_2=0)$,  where the asymptotic behaviour of the
spacetime is AdS. On the other hand, from solutions (\ref{fr2un})
and (\ref{gttexp2}) we see that for $k=\pm1$, as
$r\rightarrow\infty$, we have $\textbf{g}^{rr}
=-\textbf{g}_{tt}=k+\Lambda r^2/3$, which confirms that the
spacetime is asymptotically AdS. A close look on solution
(\ref{gttexp2}) shows that an extra term $c_0/r$ is added to the
metric, compared to the topological black hole in Einstein
gravity. This leading order term incorporates the effects of the
mimetic field on the spacetime geometry far from the horizon.
Calculating the curvature scalars of this spacetime indicates that
\begin{eqnarray}
\lim_{r\longrightarrow 0^{+}}R &=&\infty ,  \label{Rorigin} \\
\lim_{r\longrightarrow 0^{+}}R_{\mu \nu \rho \sigma }R^{\mu \nu
\rho \sigma } &=&\infty,  \label{RRorigin}
\end{eqnarray}
while, for the asymptotic region where $r\rightarrow \infty$, the
invariants of the spacetime are obtained as
\begin{eqnarray}
\lim_{r\longrightarrow \infty }R &=&-4\Lambda ,  \label{Rinf} \\
\lim_{r\longrightarrow \infty }R_{\mu \nu \rho \sigma }R^{\mu \nu
\rho \sigma } &=&\frac{8}{3}\Lambda ^{2} \, . \label{RRinf}
\end{eqnarray}

\begin{figure}[h]
  \centering
  \subfigure[$k=1$]{
    \includegraphics[width=0.45\linewidth]{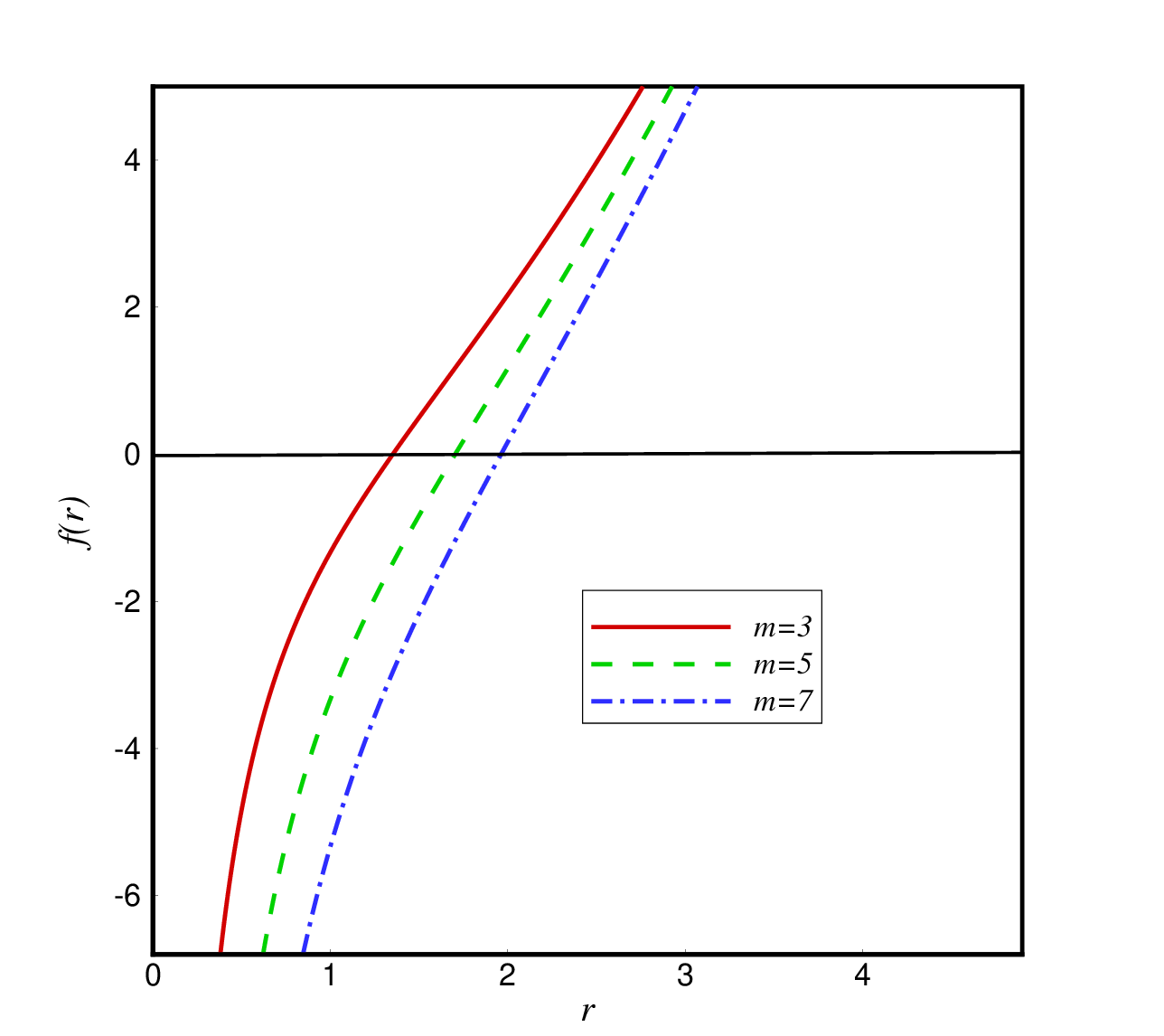}
  }
  \subfigure[$k=-1$]{
    \includegraphics[width=0.45\linewidth]{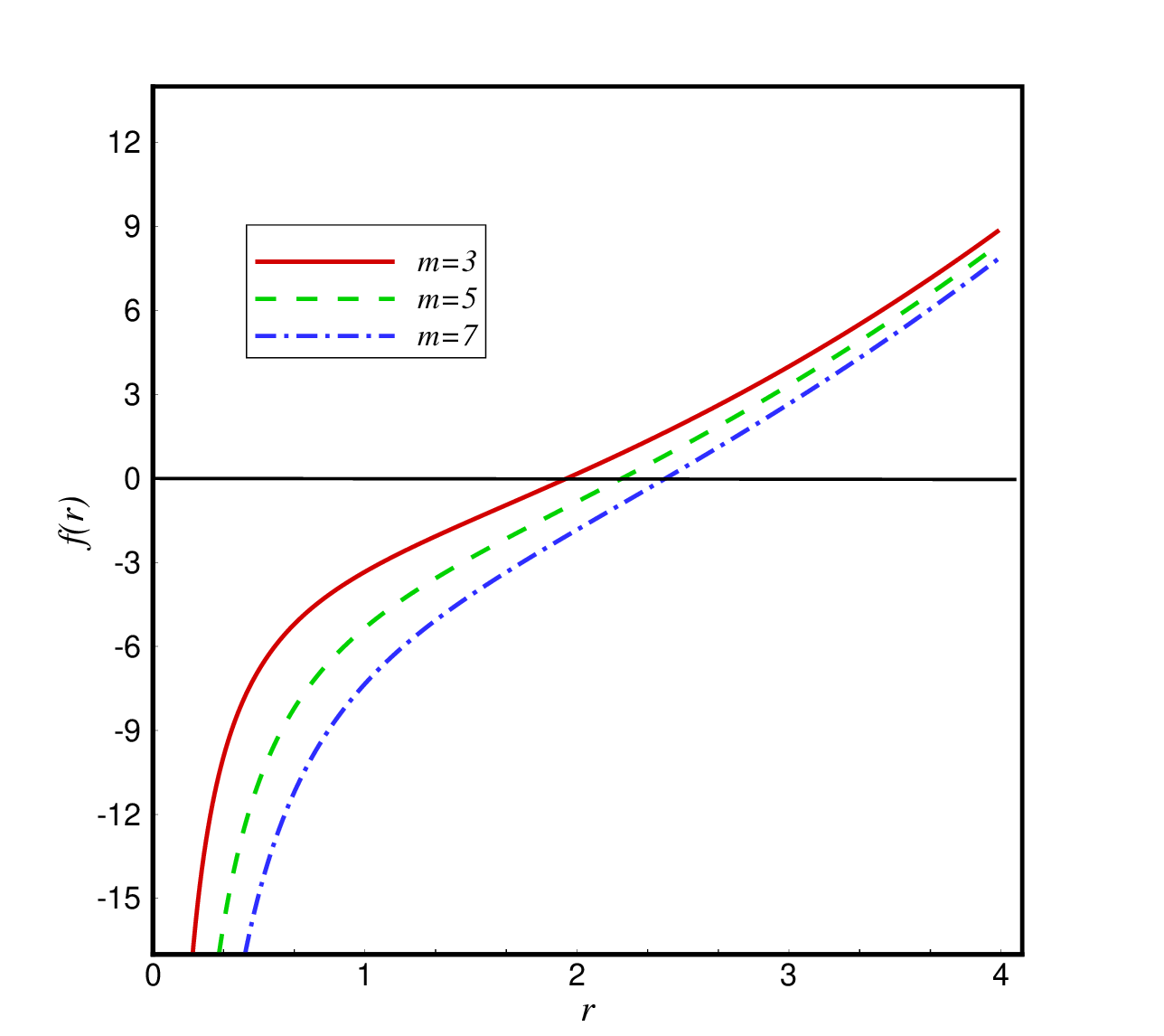}
  }
  \caption{The behavior of $f(r)$ for a topological black hole in
  the presence of a
  constant potential with $\Lambda=2$.}
  \label{Fig1,3}
\end{figure}

\begin{figure}[h]
  \centering
  \subfigure[$m=3$, $c_0=1$ and $k=1$.]{
    \includegraphics[width=0.45\linewidth]{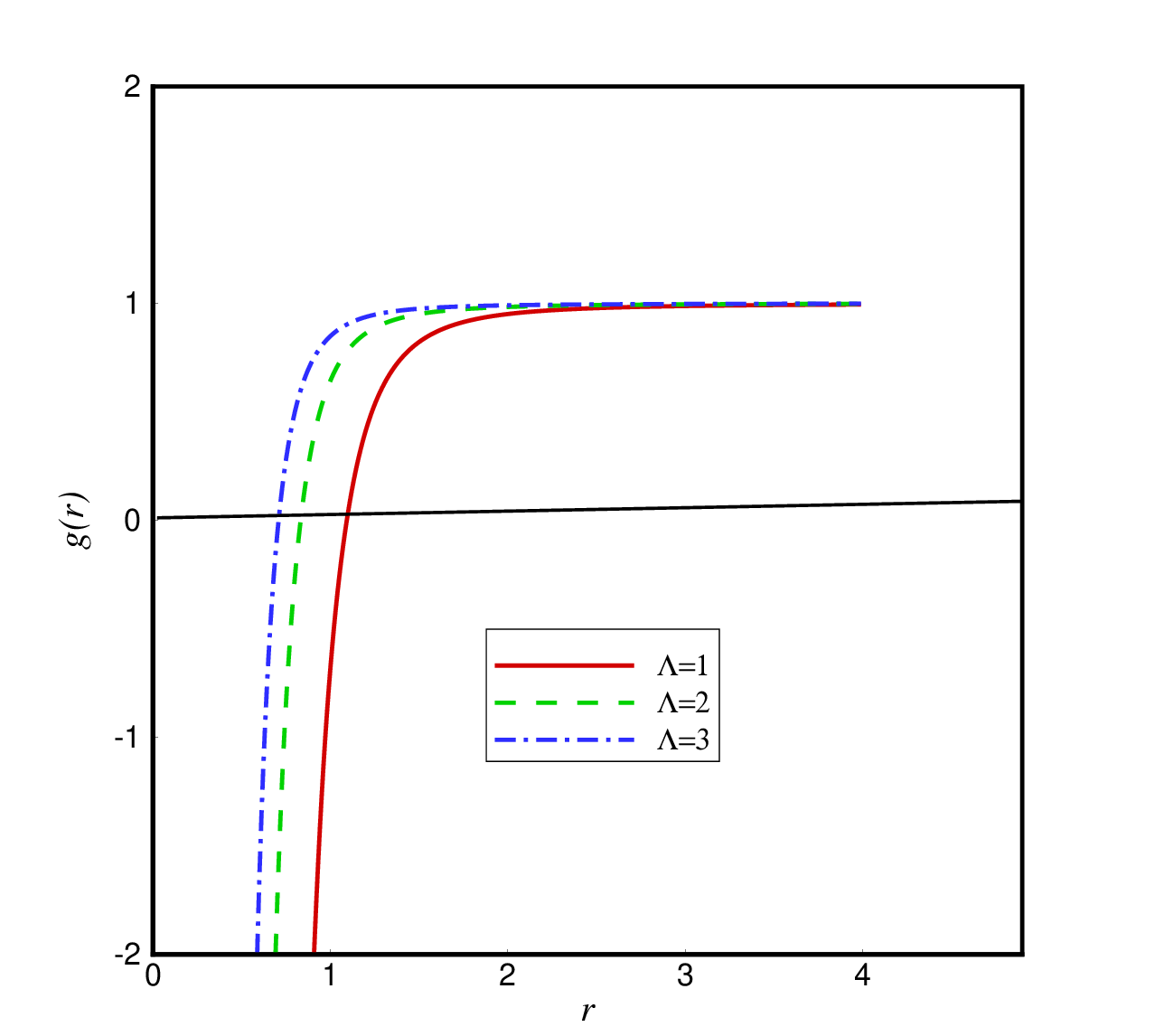}
  }
  \subfigure[$m=2$, $c_0=0.5$ and
$k=-1$]{
    \includegraphics[width=0.45\linewidth]{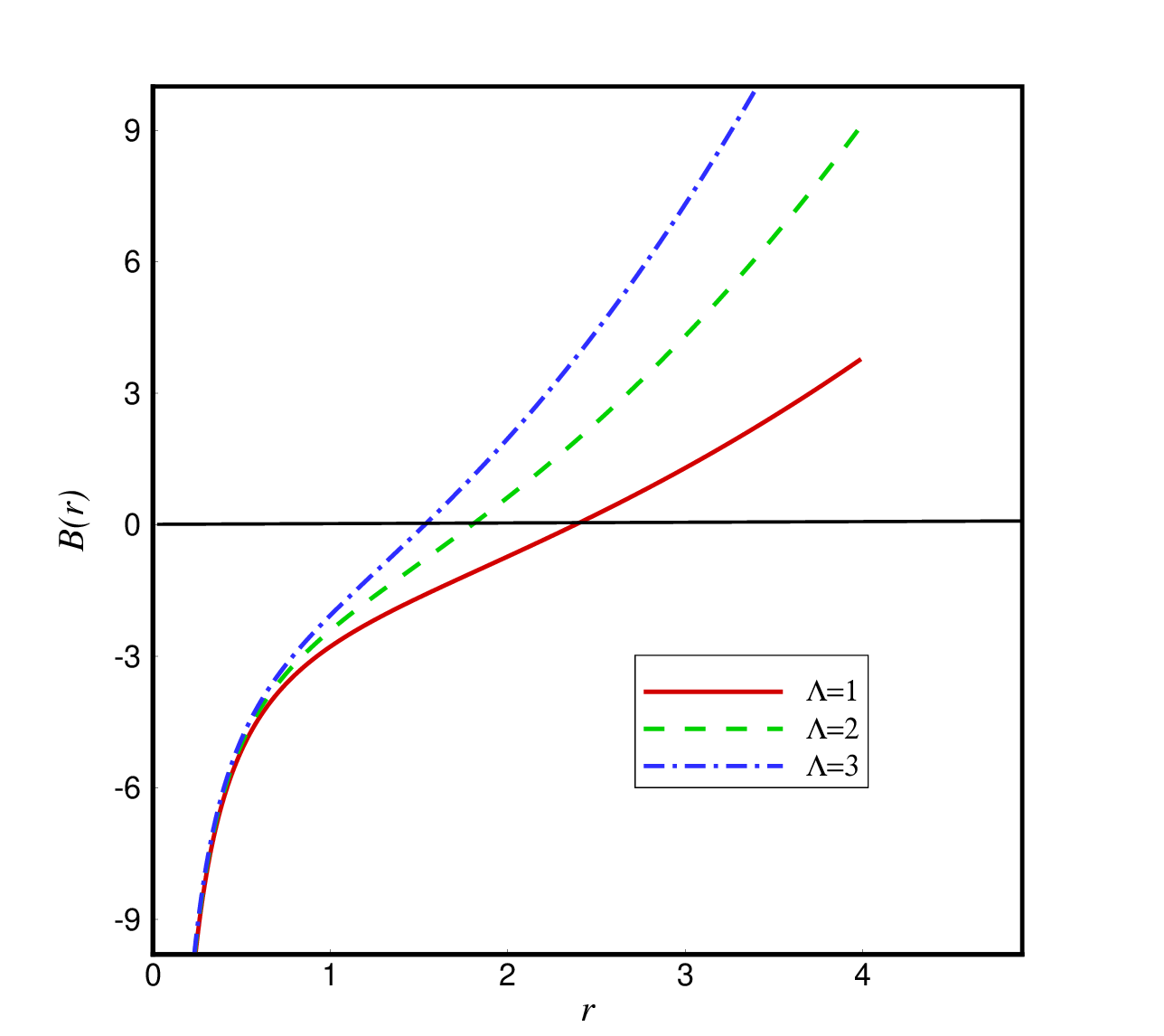}
  }
  \caption{The asymptotic behavior of the metric functions $g(r)$ (left) and $B(r)=f(r)g^2(r)$ (right) for a
topological black hole with different $\Lambda$.}
  \label{Fig2,2a}
\end{figure}

The behaviour of the metric functions for topological black holes
in mimetic gravity are shown in Figs. \ref{Fig1,3} and
\ref{Fig2,2a}. From Fig. \ref{Fig1,3}  we see that the black hole has
one horizon and the radius of this horizon increases with
increasing the mass parameter $m$. On the other hand, Fig.
(\ref{Fig2,2a}a) indicates that $g(r)\rightarrow-\infty$ for small
$r$ goes to unity for large values of $r$. This is an expected
result, since at large distance, we expect the effects of the
mimetic field to disappear. Fig. (\ref{Fig2,2a}b) also shows
that a black hole has one infinite redshift surface which is the
root of $B(r)=0$ and its radius decreases with increasing
$\Lambda$.
\section{charged black holes} \label{charge}
In this section we would like to consider the possible solutions
for charged topological black holes in the context of mimetic
gravity.
\subsection{Solution with $V(\phi)=0$}
\begin{figure}[h]
  \centering
  \subfigure[$k=1$]{
    \includegraphics[width=0.45\linewidth]{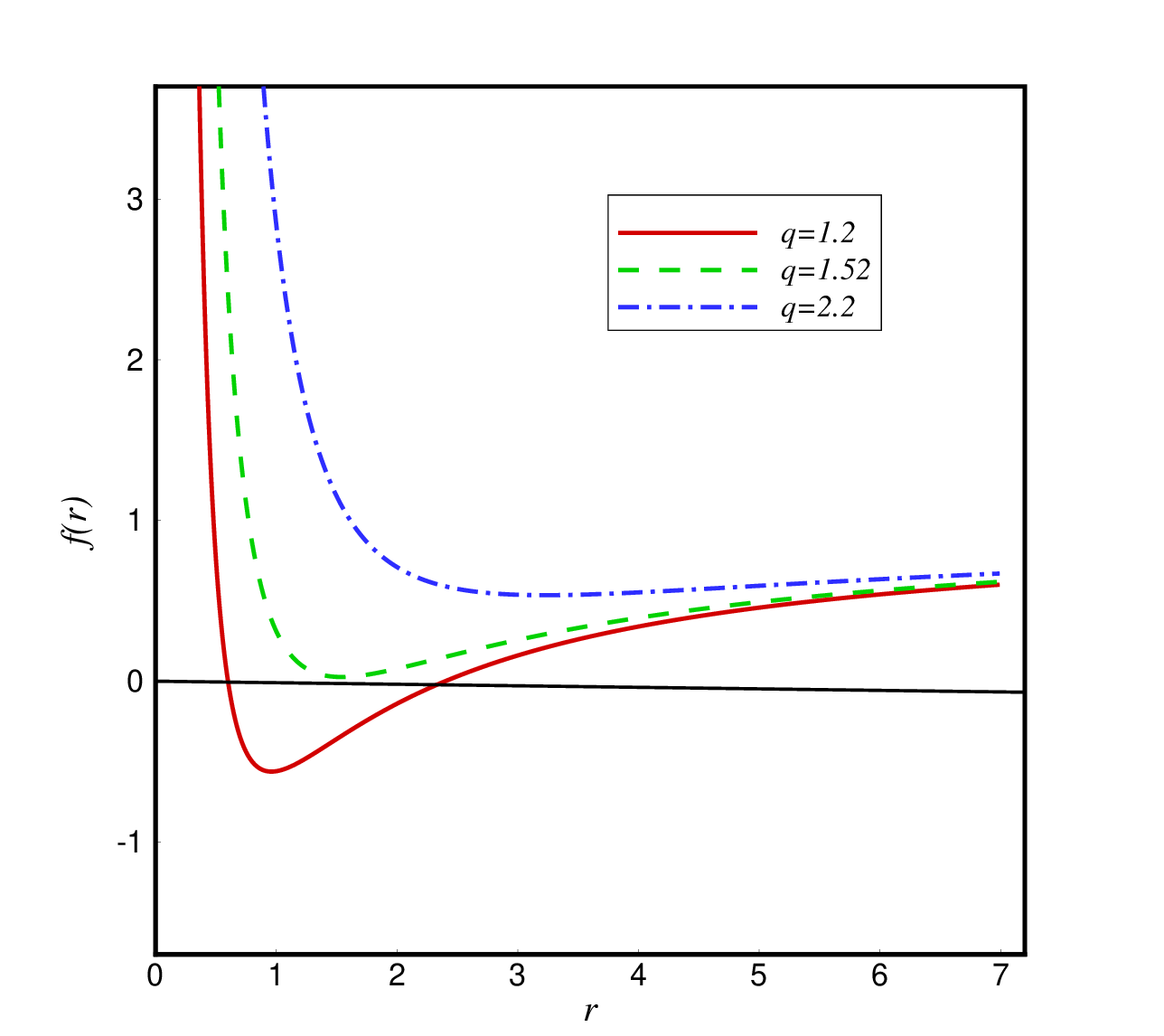}
  }
  \subfigure[$k=-1$]{
    \includegraphics[width=0.45\linewidth]{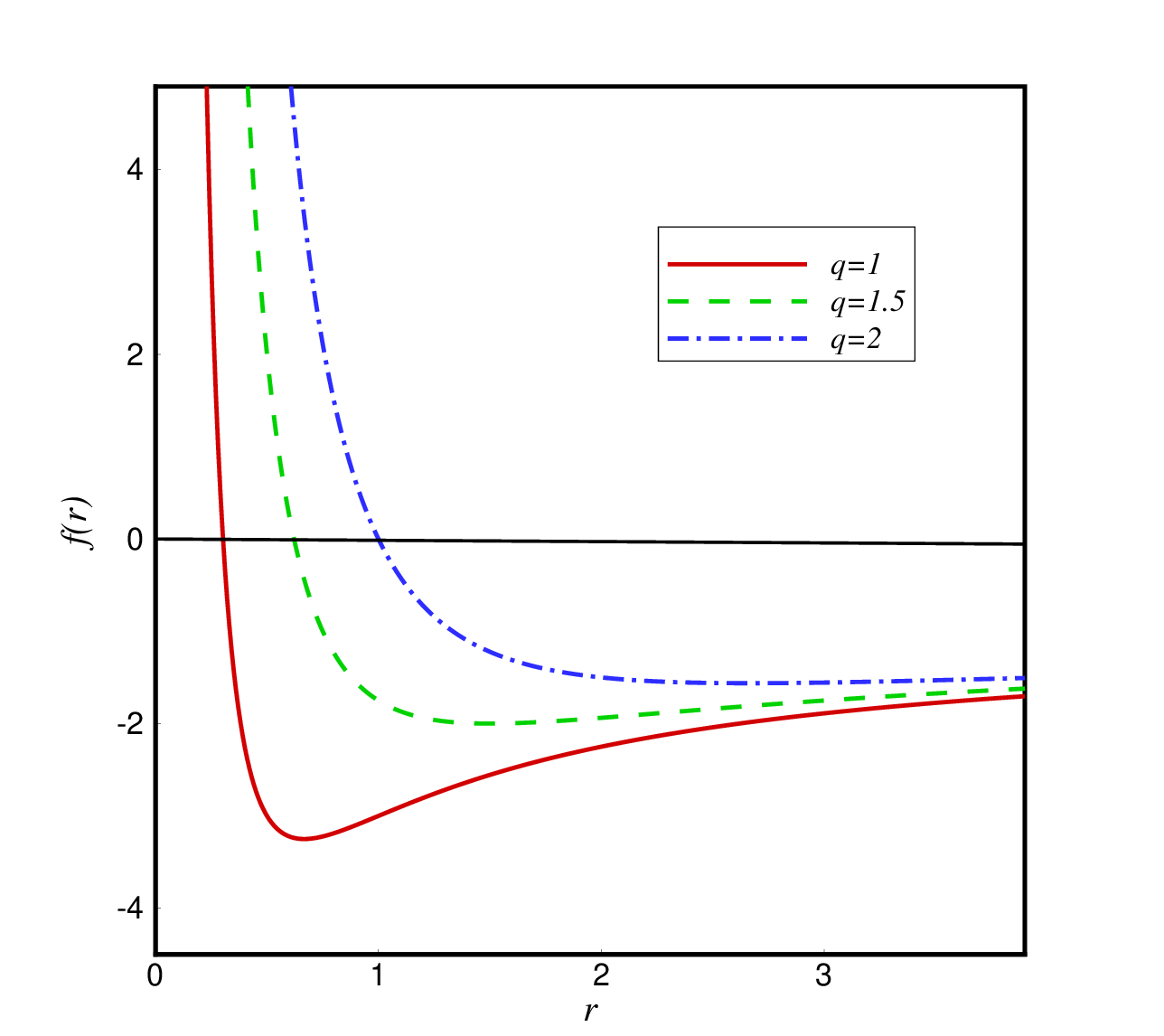}
  }
  \caption{The behavior of $f(r)$ for a charged topological black
  hole with $m=3$ and different $q$.}
  \label{Fig4ac}
\end{figure}

To have a better insight on the nature of the solutions we first
consider the case with zero potential. In this case Eqs.
(\ref{tt})-(\ref{pp}) have the following solutions
\begin{eqnarray}\label{frch}
f(r)&=&k-\frac{m}{r}+\frac{q^2}{r^2}, \  \quad\quad\quad  \quad\quad\quad {\rm for} \ k=0,\pm 1\\
g(r)&=&1+B_0 \frac{m^2 r^2+4q^2 r m -8 q^4}{\sqrt{mr-q^2}},
\quad\!\!{\rm
for}\quad k=0,\nonumber\\
g(r)&=& 1+A_0 \Bigg{\{}\ln
\left(\frac{r}{r_0}+\frac{\sqrt{r^2-k(mr-q^2)}}{r_0}-\frac {km}{2
r_0} \right)- \frac{2(m^2 r-2 kq^2 r -mq^2)}{(m^2-4 kq^2)
\sqrt{r^2-k(mr-q^2)}} \Bigg{\}}, \ {\rm for} \ k=\pm 1, \nonumber\\
 \label{grch}
\end{eqnarray}
where $B_0$, $A_0$ and $r_0$ are constants of integration. In this
case we have a black hole solution for $k=1$. For $k=0,-1$, however,
we have no black hole solution and one may encounter with a naked
singularity covered by a cosmological horizon. Indeed, in these
cases the metric function changes its sign, $f(r)<0$, for $r>r_c$
as one can see in Fig. (\ref{Fig4ac}b). For $k=1$, however, we
have two horizons located at,
\begin{eqnarray}
r_{\pm}= \frac{1}{2} \left(m\pm \sqrt{m^2-4 q^2}\right).
\end{eqnarray}
Thus, for spherical topology, we have a black hole with an inner
and an outer horizon for $m>2q$, an extremal black hole for $m=2q$,
and a naked singularity for $m<2q$ (see  Fig. \ref{Fig4ac}a ).

Next, we study the behaviour of the metric function $B(r)=f(r)
g^2(r)$ for large values of $r$. It is a matter of calculation to
show that for $k=\pm1$, at far distance,
\begin{eqnarray}\label{g00cexp1}
B(r)\approx &=& A_1+A_2 \ln \left(\frac{r}{r_0}\right)+ A_3
\frac{\ln r}{r}+ \frac{A_4}{r}+...,
\end{eqnarray}
while for $k=0$, $B(r)\approx B_1 r^2 + B_2 r +...$ which diverges
as $r\rightarrow \infty$. This might be due to the presence of the
mimetic field $\phi$. Here $A_i$ and $B_i$ are constants which are
functions of $m$, $r_0$, $q$, $B_0$ and $A_0$. In conclusion, in
the absence of a potential, we have charged black hole solutions
only for spherical horizon topology ($k=1$), and in other cases
($k=0,-1$) we encounter a naked singularitiy with a cosmological
horizon. In all cases the spacetime is asymptotically neither flat
nor AdS due to the presence of the mimetic field.
\subsection{Solutions with $V(\phi)=-2\Lambda$}
Finally, we consider charged topological black holes in the
presence of a potential for the mimetic field. In this case Eqs.
(\ref{tt})-(\ref{pp}) admit the following solutions
\begin{eqnarray}\label{frch2}
f(r)&=&k-\frac{m}{r}+\frac{q^2}{r^2}+\frac{\Lambda  r^2}{3}, \quad\!\!{\rm for}\quad k=0,\pm 1\\
g(r)&=&1+C_0 \int{\frac{r^2 dr}{(3k r^2+\Lambda r^4 -3m
r+3q^2)^{3/2}}},  \quad\! \quad\! \quad\!\!{\rm for}\quad k=0,\pm
1 \label{gch2}
\end{eqnarray}
where $C_0$ is a constant of integration. The integral in the
expression (\ref{gch2}) cannot be done analytically, so we expand
the integrand for large $r$ limit. It is a matter of calculations
to show that
\begin{eqnarray}\label{grchexp}
g(r)&\approx& 1+ \frac{C_0}{\Lambda^{5/2} r^5}
\Bigg{\{}\frac{9k}{10}-\frac{\Lambda r^2}{3}-\frac{3m}{4r}+\frac{9
q^2}{14 r^2}-\frac{135|k|}{56 \Lambda r^2}+...\Bigg{\}},
\end{eqnarray}
which yields
\begin{eqnarray}\label{g00cexp2}
B(r)=f(r) g^2(r)\approx k-\frac{m}{r}-\frac{ 2 C_0}{9
\sqrt{\Lambda} r}+\frac{q^2}{r^2}+ \frac{\Lambda r^2}{3}+...
\end{eqnarray}
Clearly, in the asymptotic region where $r\rightarrow \infty$, we
have  $g(r)\approx 1$ and
$\textbf{g}^{rr}=-\textbf{g}_{tt}=k+\Lambda r^2/3$, which confirms
that the asymptotic behaviour of the obtained solutions is AdS.
Comparing the above solution with the solution of Einstein gravity
in the presence of a cosmological constant, here, we have an
additional term $-\frac{ 2 C_0}{9 \sqrt{\Lambda} r}$ which
incorporates the effect of the mimetic field through the constant
$C_0$. If we redefine the mass parameter such as $m'=m+2C_0/(9
\sqrt{\Lambda})$, then the above solution reduces to
\begin{eqnarray}\label{g00cexp3}
B(r)\approx k-\frac{m'}{r}+\frac{q^2}{r^2}+ \frac{\Lambda
r^2}{3}+...
\end{eqnarray}
Let us have a look at the electric field of the spacetime. From Eq.
(\ref{Er}), we have
\begin{equation}\label{Er2}
E(r)=\frac{q}{r^2}g(r)\approx  \frac{q}{r^2}\Bigg{\{} 1+ \frac{
C_0}{\Lambda^{5/2} r^5}\left[ \frac{9k}{10}-\frac{\Lambda
r^2}{3}-\frac{3m}{4r}+\frac{9 q^2}{14 r^2}-\frac{135|k|}{56
\Lambda r^2}+...\right]\Bigg{\}}
\end{equation}
Thus, far from the black hole, the electric field behaves as
$E(r)\sim q/r^2$, since in this regime we have $g(r)\approx 1$.
Again the Ricci and Kretschmann invariants diverge at $r=0$, they
are finite for $r\neq0$ and go to $-4\Lambda$ and $4\Lambda^2/3$,
respectively, for $r\rightarrow \infty$. Thus, we confirm that the
essential singularity is at $r=0$.

\begin{figure}[h]
  \centering
  \subfigure[$k=1$]{
    \includegraphics[width=0.45\linewidth]{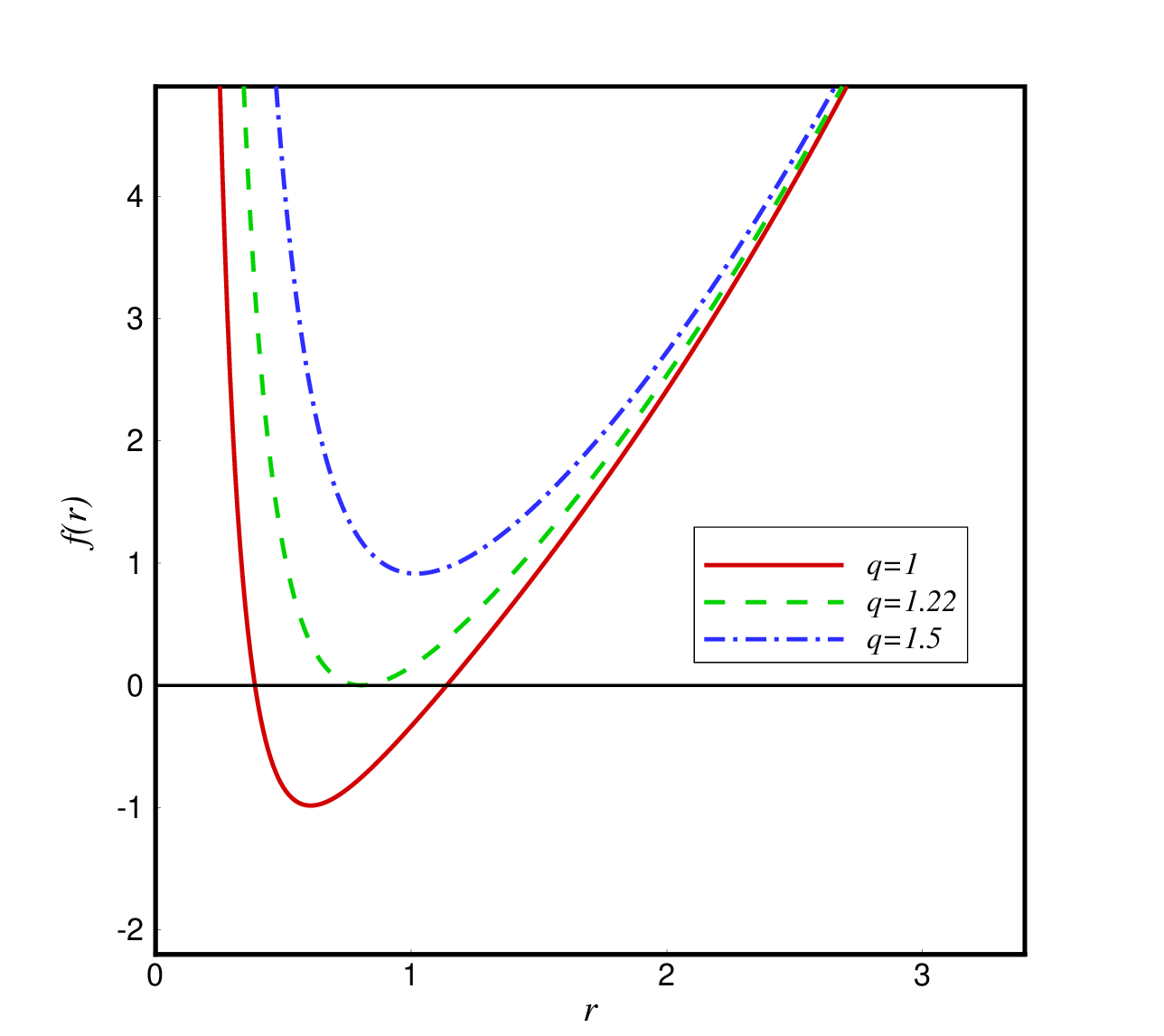}
  }
  \subfigure[$k=-1$]{
    \includegraphics[width=0.45\linewidth]{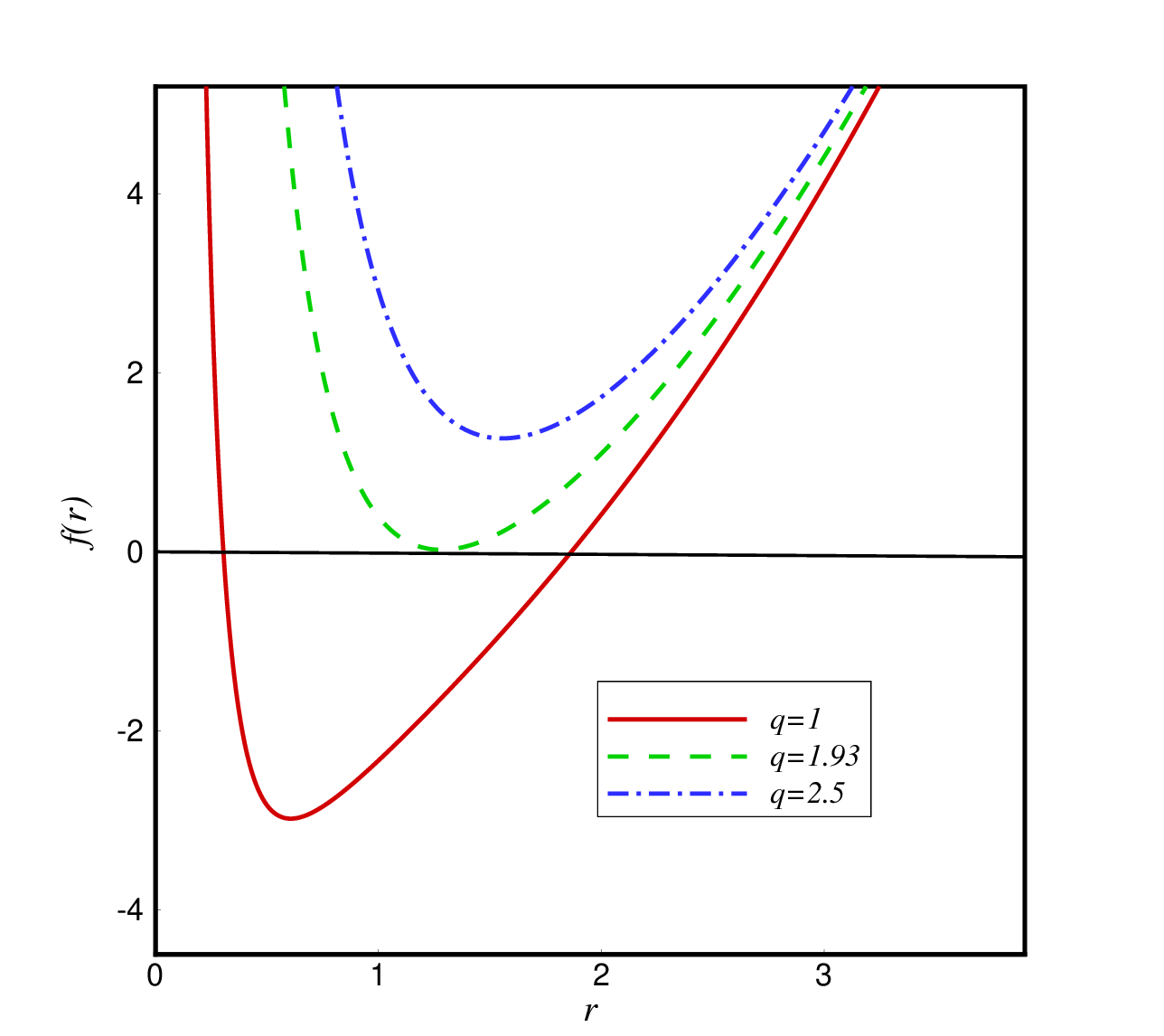}
  }
  \caption{The behavior of $f(r)$ for a charged topological black
  hole in the presence of a constant potential
with $\Lambda=2$, $m=3$.}
  \label{Fig44b}
\end{figure}
\begin{figure}[h]
  \centering
  \subfigure[$m=2$, $\Lambda=2$ and
$k=1$]{
    \includegraphics[width=0.45\linewidth]{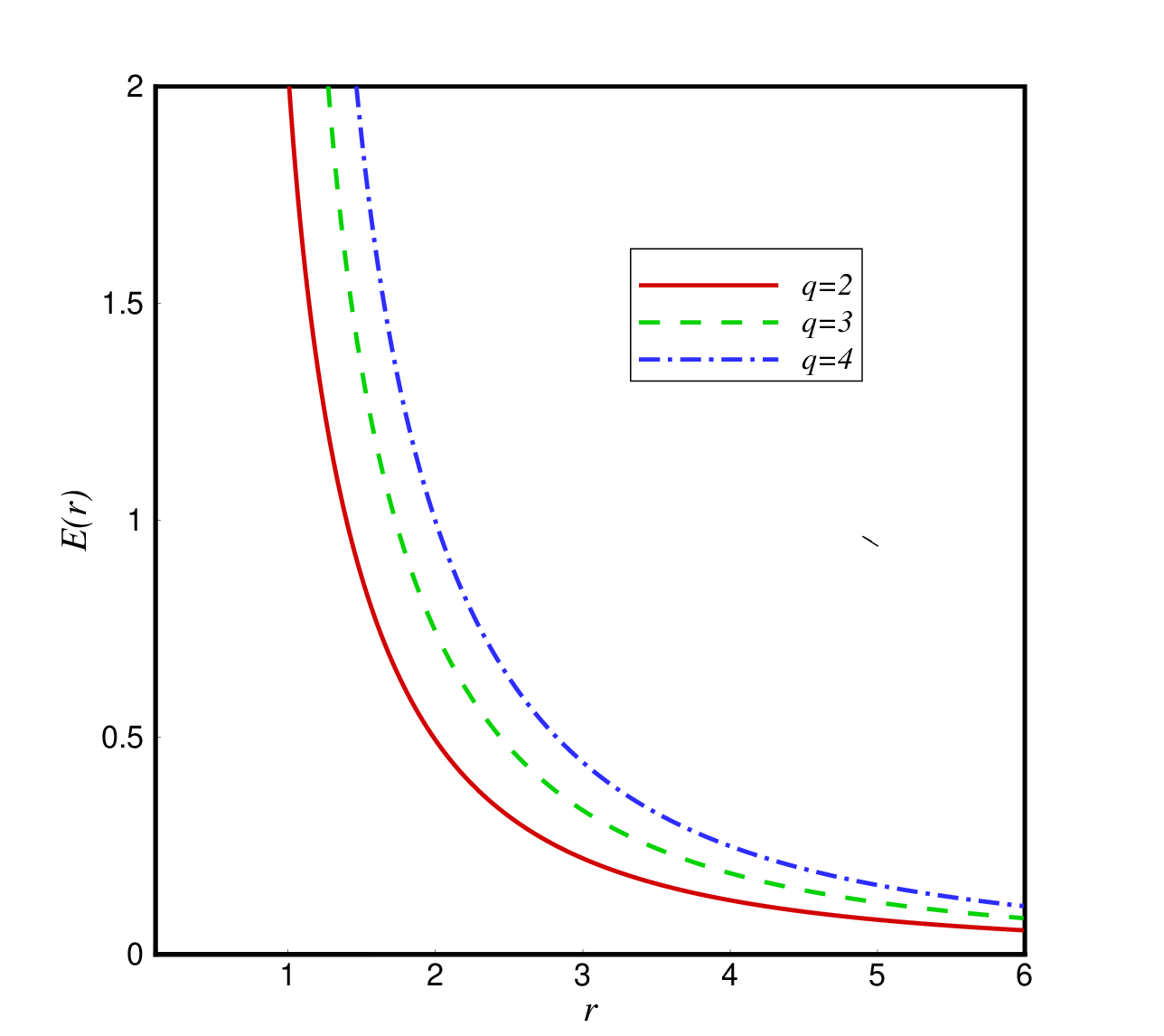}
  }
  \subfigure[$\Lambda=1$, $q=1.5$, and $k=0$]{
    \includegraphics[width=0.45\linewidth]{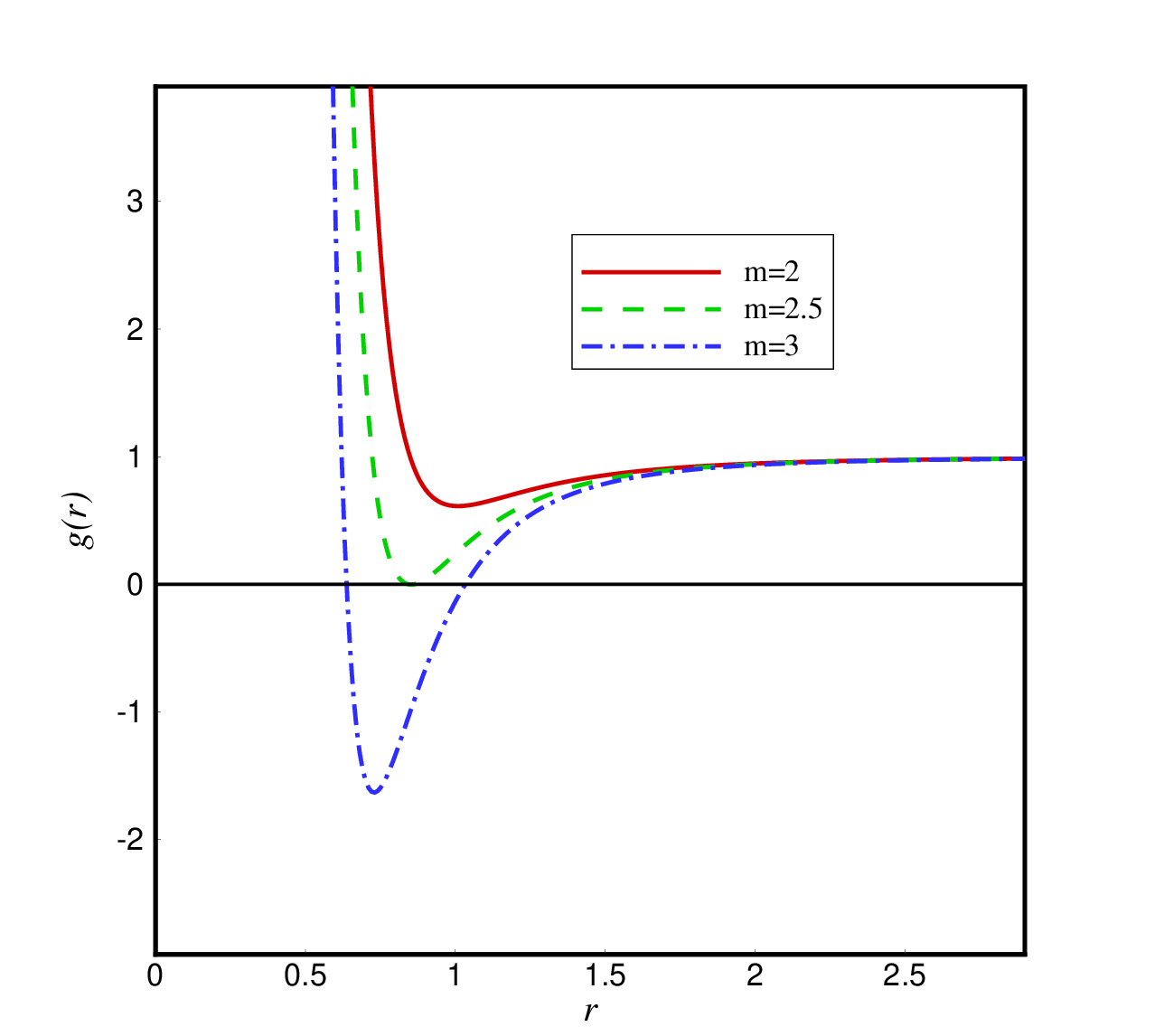}
  }
  \caption{The behavior of the electric field $E(r)$ (left) and $g(r)$ (right) for a charged topological black
  hole in the presence of a constant potential with $C_0=1$.}
  \label{Fig56}
\end{figure}
The behavior of the metric functions and the electric field for
charged topological black holes in mimetic gravity are plotted in
Figs. \ref{Fig44b} and \ref{Fig56}. From Fig. \ref{Fig44b}, we
observe that our solutions can represent, depending on the metric
parameters, black holes with an inner and an outer horizon, an
extremal black hole, or naked singularity. Similar behaviour is also
seen for the cases with $k=0$. In Fig. (\ref{Fig56}a), we also
plotted the behaviour of the electric field, where it is observed
that, for small $r$ the electric field diverges, while it goes to
zero for large values of $r$, as one expects. Finally, from Fig.
(\ref{Fig56}b), we see that $g(r)$ diverges for small values of
$r$ and goes to unity for large values of $r$.
\newpage
\section{Geodesics motion around mimetic black holes} \label{geodesic}
In this section we study the geodesic motion of massless and
massive particles in the spacetimes of mimetic black holes. For
symmetry reasons we can choose $\theta$ to be constant. In the
case $k=1$ we will choose the equatorial plane
$\theta=\frac{\pi}{2}$ and in the case $k=-1$ we will choose
$\theta=\ln(1+\sqrt{2})$ so that $\sinh(\theta)=1$. Then in all
three cases $k=1,0,-1$ we will have $g_{\varphi\varphi}=r^2$.

We derive the equations of motion using the Hamilton-Jacobi
formalism. To solve the Hamilton-Jacobi equation
\begin{equation}
 \half g^{\mu\nu} \frac{\partial S}{\partial x^\mu} \frac{\partial S}{\partial x^\nu}  + \frac{\partial S}{\partial \tau}
 =0,
 \label{eqn:hjeq}
\end{equation}
we use the following ansatz for the action
\begin{equation}
 S= \half\delta\tau - Et + L\varphi + S_r(r),
\end{equation}
where $\tau$ is an affine parameter along the geodesics. For
massive particles we have $\delta=1$ and for massless particles
$\delta=0$. $E$ is the energy and $L$ is the angular momentum of
the test particle. With the above ansatz and the metric
\eqref{metric} the Hamilton-Jacobi equation becomes
\begin{equation}
 \delta-\frac{E^2}{f(r)g(r)^2}+\frac{L^2}{r^2}+f(r)\left(\frac{\partial S_r}{\partial r}\right)^2 = 0,
\end{equation}
and therefore
\begin{equation}
 \left(\frac{\partial S_r}{\partial r}\right) = \sqrt{ -\frac{\delta}{f(r)} -\frac{L^2}{f(r)r^2} +\frac{E^2}{f(r)^2g(r)^2} } \  .
\end{equation}
We can derive the equations of motion by varying the action with respect to the constants of motion
\begin{align}
 \left( \frac{\dd r}{\dd \varphi} \right) ^2 &= \frac{r^4 E^2}{L^2 g(r)^2}-  r^2 f(r) \left( \frac{\delta r^2}{L^2} -1\right) \, , \label{eqn:rphi}\\
 \left( \frac{\dd r}{\dd t} \right) ^2 &= f(r)^2 g(r)^2-\frac{ f(r)^3 g(r)^4}{E^2} \left(\delta +\frac{L^2 }{r^2}\right) \, . \label{eqn:rt}
\end{align}
From equation \eqref{eqn:rphi} we can define an effective potential $V_{\rm eff}$ by
\begin{equation}
 \left( \frac{\dd r}{\dd \phi} \right) ^2 = \frac{r^4}{L^2 g(r)^2} \left( E^2 - V_{\rm eff}\right)
\end{equation}
which yields
\begin{equation}
 V_{\rm eff}= \left(\delta+\frac{L^2}{r^2}\right) f(r) g(r)^2 \, .
 \label{eqn:Veff}
\end{equation}

\subsection{Uncharged black holes with $V(\phi)=0$}

\begin{figure}[h]
  \centering
  \subfigure[~$\delta=1$]{
    \includegraphics[height=0.4\linewidth]{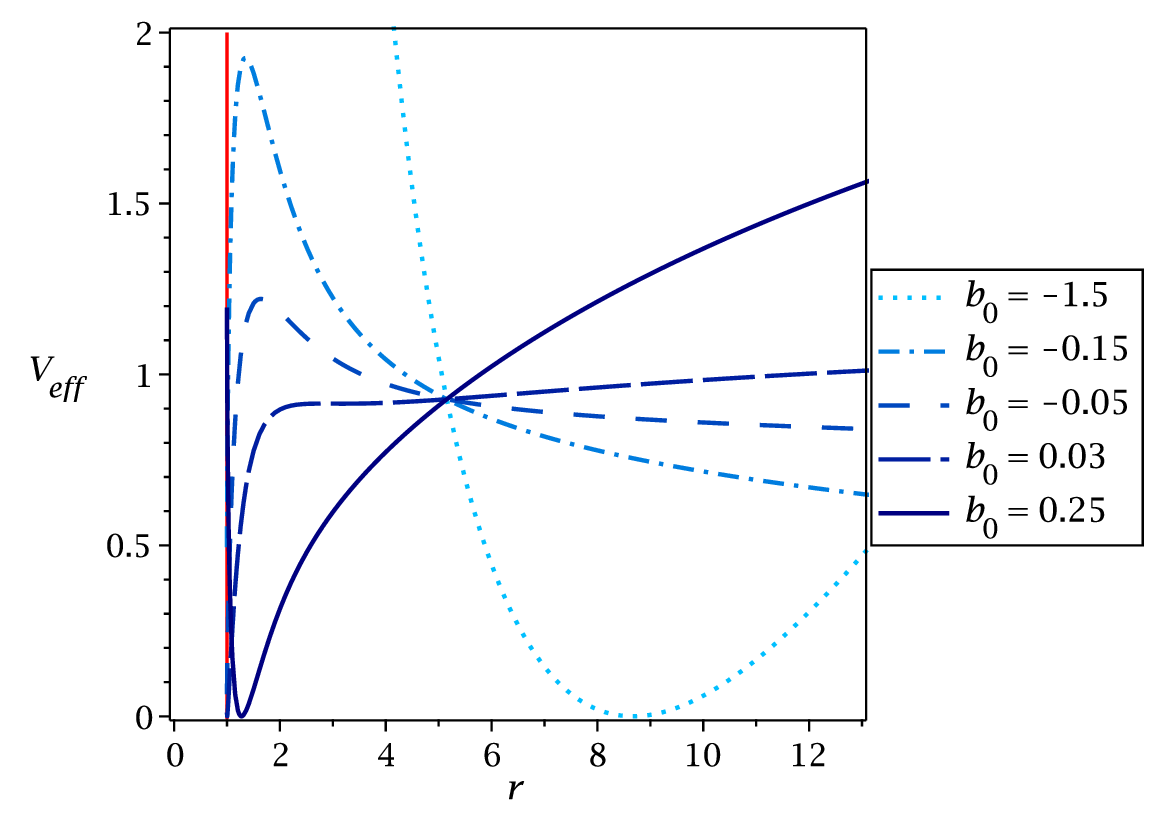}
  }
  \subfigure[~$\delta=0$]{
    \includegraphics[height=0.4\linewidth]{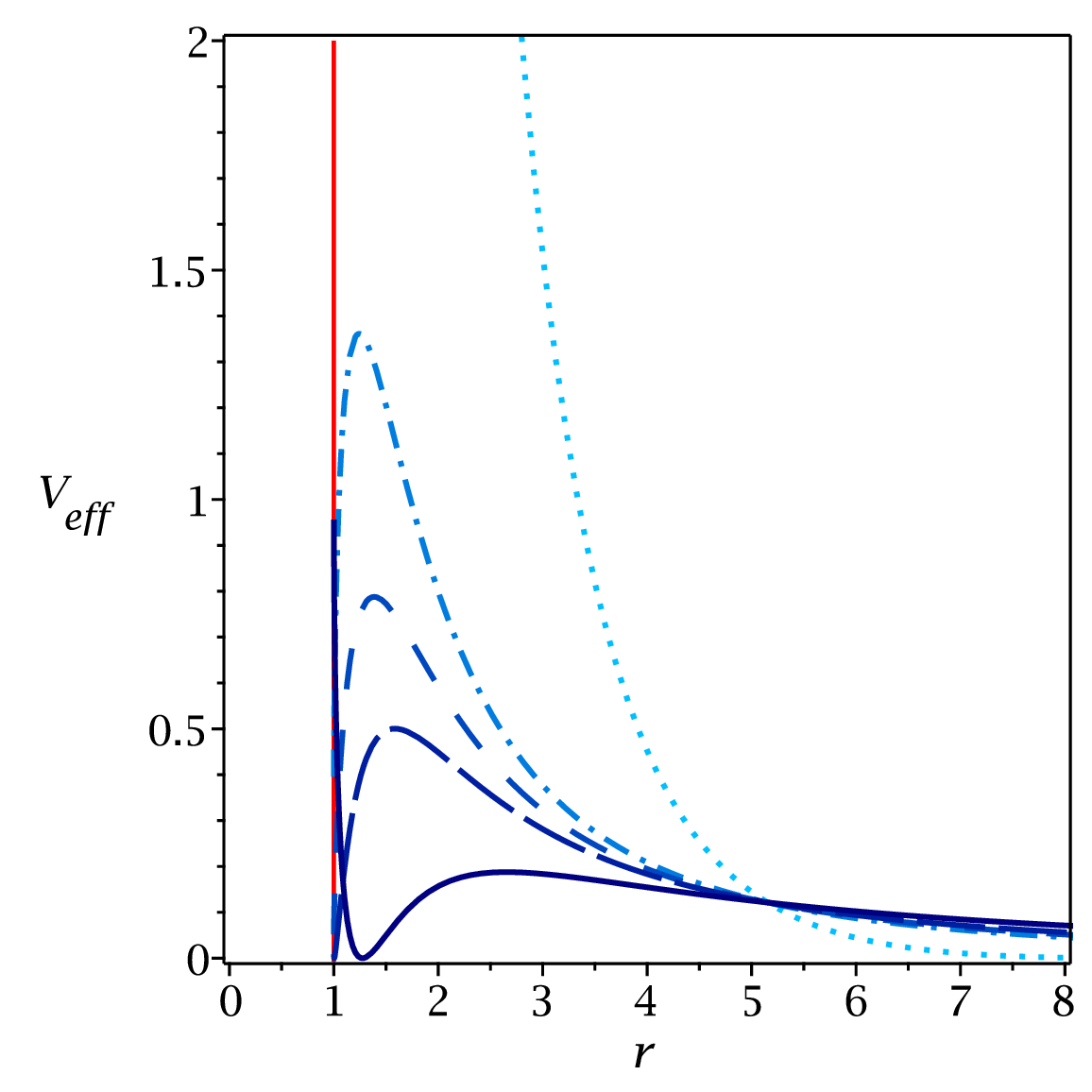}
  }
  \caption{Effective potential \eqref{eqn:Vk1} for test particles moving around an uncharged black hole with $V(\phi)=0$ and $k=1$, $m=1$, $L=2$, $r_0=1$. The blue curves represent the effective potential for different values of $b_0$ and the red vertical line indicates the position of the event horizon. Figure (a) is the effective potential for massive particles and figure (b) is the effective potential for light.}
  \label{pic:pot_uncharged_V0}
\end{figure}

First we will consider uncharged black holes with $V(\phi)=0$. We
insert the metric functions \eqref{frun} and \eqref{grun} the case
$k=1$, which describes a spherical black hole in mimetic gravity.
The $r(\phi)$ equation of motion and the effective potential are
\begin{align}
  \left( \frac{\dd r}{\dd \varphi} \right) ^2 &= \frac{E^2r^4}{L^2}  \left\{ 1+b_0 \left[-2 \left(1-\frac {m}{r}\right)^{-1/2}+\ln
\left( \frac{r}{r_0}\left(+\sqrt{ 1-{\frac {m}{r}}}\right)-\frac {m}{2r_0} \right)\right] \right\} ^{-2} \nonumber \\
&- r^2\left(1-\frac{m}{r}\right)\left(\frac{\delta r^2}{L^2} -1\right) \, , \label{eqn:rphik1}\\
  V_{\rm eff} &= \left(\delta+\frac{L^2}{r^2}\right)\left(1-\frac{m}{r}\right)  \left\{ 1+b_0 \left[-2 \left(1-\frac {m}{r}\right)^{-1/2}+\ln
\left( \frac{r}{r_0}\left(+\sqrt{ 1-{\frac {m}{r}}}\right)-\frac {m}{2r_0} \right)\right] \right\}^2\, . \label{eqn:Vk1}
\end{align}
Apparently equation \eqref{eqn:rphik1} can only be solved with
numerical methods. However, we can analyse the effective potential
\eqref{eqn:Vk1} to study the behaviour of the geodesics. Figure
\ref{pic:pot_uncharged_V0} shows some plots of the effective
potential \eqref{eqn:Vk1} for different values of $b_0$. We see
that for every $b_0$ all potentials meet at the same point $(r_p,
V_{\rm eff}(r_p))$, which can be calculated by
\begin{equation}
 r_0 \exp\left(\frac{2}{\sqrt{1-\frac{m}{r_p}}}\right) = r_p\sqrt{1-\frac{m}{r_p}}-\frac{m}{2}+r_p \, .
\end{equation}
Furthermore, if $g(r_c)=0$, then $V_{\rm eff}(r_c)=\frac{\dd
V_{\rm eff}}{\dd r}(r_c)=0$ and there is a stable circular orbit
with zero energy. The position $r_c$ of the circular orbit can be
calculated by
\begin{equation}
  r_0 \exp\left(\frac{2b_0-\sqrt{1-\frac{m}{r_p}}}{b_0\sqrt{1-\frac{m}{r_p}}}\right) = r_p\sqrt{1-\frac{m}{r_p}}-\frac{m}{2}+r_p \, .
\end{equation}

Due to the logarithmic term, the effective potential diverges for
$r\rightarrow\infty$ in the case $\delta=1$. This means that all
geodesics for massive particles are bound orbits, as long as
$b_0\neq 0$. For $\delta=0$ the effective potential approaches
$V_{\rm eff}=0$ for $r\rightarrow\infty$, so that light can escape
the black hole.

The turning points of the orbits are the zeros of $\left(
\frac{\dd r}{\dd \varphi} \right) ^2$. Geodesic motion is possible
for $\left( \frac{\dd r}{\dd \varphi} \right) ^2\geq 0$, which
means $E^2\geq V_{\rm eff}$. We find the following orbit
configurations for massive particles:
\begin{enumerate}
 \item A single turning point $r_1>r_h$ exists and $\left( \frac{\dd r}{\dd \varphi} \right) ^2\geq 0$ for $r_1 \geq r \geq 0$. The corresponding orbit is a terminating bound orbit, which ends in the singularity at $r=0$.
 \item Two turning points $r_2 \geq r_1 >r_h$ exist and $\left( \frac{\dd r}{\dd \varphi} \right) ^2\geq 0$ for $r_2 \geq r \geq r_1$. The corresponding orbit is a bound orbit.  This configuration does not exist in the Schwarzschild spacetime.
 \item Three turning points $r_3 >r_2 > r_1 >r_h$ exist and  $\left( \frac{\dd r}{\dd \varphi} \right) ^2\geq 0$ for $r_3 \geq r \geq r_2$ and  $r_1 \geq r \geq 0$. There is a bound orbit with $r_3\geq r \geq r_2$ and a terminating bound orbit with $r_1 \geq r \geq 0$ , which ends in the singularity.
\end{enumerate}
In the case of massless particles the following orbits occur:
\begin{enumerate}
 \item There is no turning point and $\left( \frac{\dd r}{\dd \varphi} \right) ^2\geq 0$ for $r\geq 0$. The light rays coming from infinity fall into the singularity (terminating escape orbit).
 \item Two turning points $r_2 \geq r_1 >r_h$ exist and $\left( \frac{\dd r}{\dd \varphi} \right) ^2\geq 0$ for $r_1 \geq r$ and $r \geq r_2$. A terminating bound orbit and an escape orbit exist.
 \item Three turning points $r_3 >r_2 > r_1 >r_h$ exist and  $\left( \frac{\dd r}{\dd \varphi} \right) ^2\geq 0$ for $r_2 \geq r \geq r_1$ and  $r\geq r_3$. There is a bound orbit and an escape orbit. This configuration does not exist in the Schwarzschild spacetime.
\end{enumerate}

If the parameter $|b_0|$ is sufficiently small, the geodesics
close to the black hole in mimetic gravity behave similar to those
in the Schwarzschild spacetime. However for large $r$ the
logarithmic term dominates, so that in contrast to General
Relativity massive particles cannot escape the mimetic black hole.
As the parameter $|b_0|$ grows, the effective potential close to
the black hole is no longer similar to Schwarzschild and a minimum
at $E=0$ appears in the effective potential. Then there are stable
circular orbits with zero energy. Moreover, massless particles can
move on a stable bound orbit, which is not possible in the
Schwarzschild spacetime.

Looking at the sky an observer will see a dark region, the so
called shadow, around a black hole. The shadow is a projection of
the photon sphere \cite{Claudel:2000yi}, which marks the boundary
between light rays escaping the black hole and light rays falling
into the black hole. For small $|b_0|$ there is an unstable
circular orbit for photons, which corresponds to the radius of the
photon sphere. Here the shadow is very similar to the
Schwarzschild spacetime. If we increase $|b_0|$, the size of the
shadow will change.

For negative $b_0$ there is still an unstable circular orbit for
photons which will be closer to the black hole in comparison to
the Schwarzschild case.

For positive $b_0$, the situation is more complex. At first, for
small $b_0$ we have an unstable circular orbit for photons, which
will be further away from the black hole than in the Schwarzschild
case. For increasing $b_0$ there is a stable circular orbit for
photons, but additionally there is an unstable circular orbit for
larger $r$ (see figure \ref{pic:pot_uncharged_V0} of the effective
potential). This unstable photon orbit is the photon sphere, which
is at a larger distance than in the Schwarzschild case.

Now if we increase $b_0$ further, we have to take into account,
that the effective potential at the event horizon is not zero, as
in the Schwarzschild case. At some point the energy value of the
effective potential at the horizon will be larger that the energy
at the unstable circular photon orbit. This means that light rays
can get arbitrarily close to the black hole and still beeing
reflected at the potential barrier (of course at a certain energy
they will fall beyond the horizon). In this case the photon sphere
has the same size as the horizon.

\subsection{Charged black holes with $V(\phi)=-2\Lambda$}

\begin{figure}[h]
  \centering
  \subfigure[~$\delta=1$]{
    \includegraphics[height=0.4\linewidth]{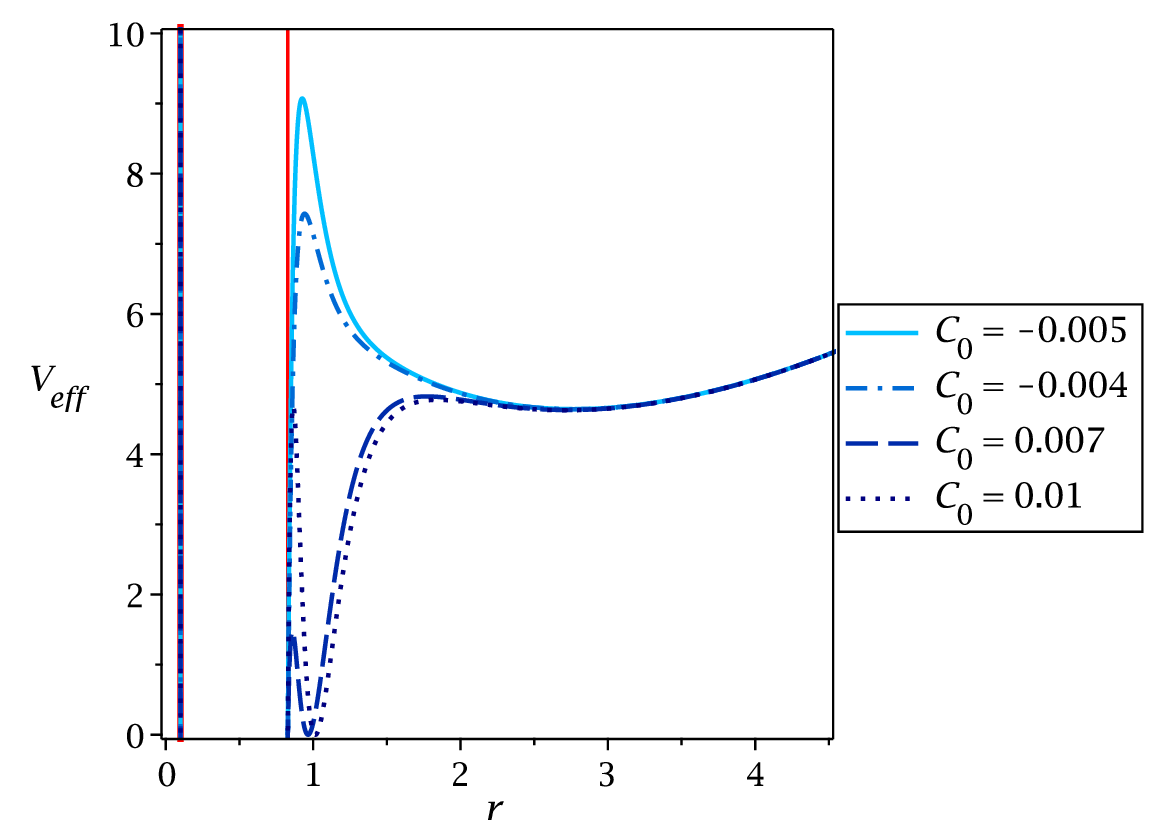}
  }
  \subfigure[~$\delta=0$]{
    \includegraphics[height=0.4\linewidth]{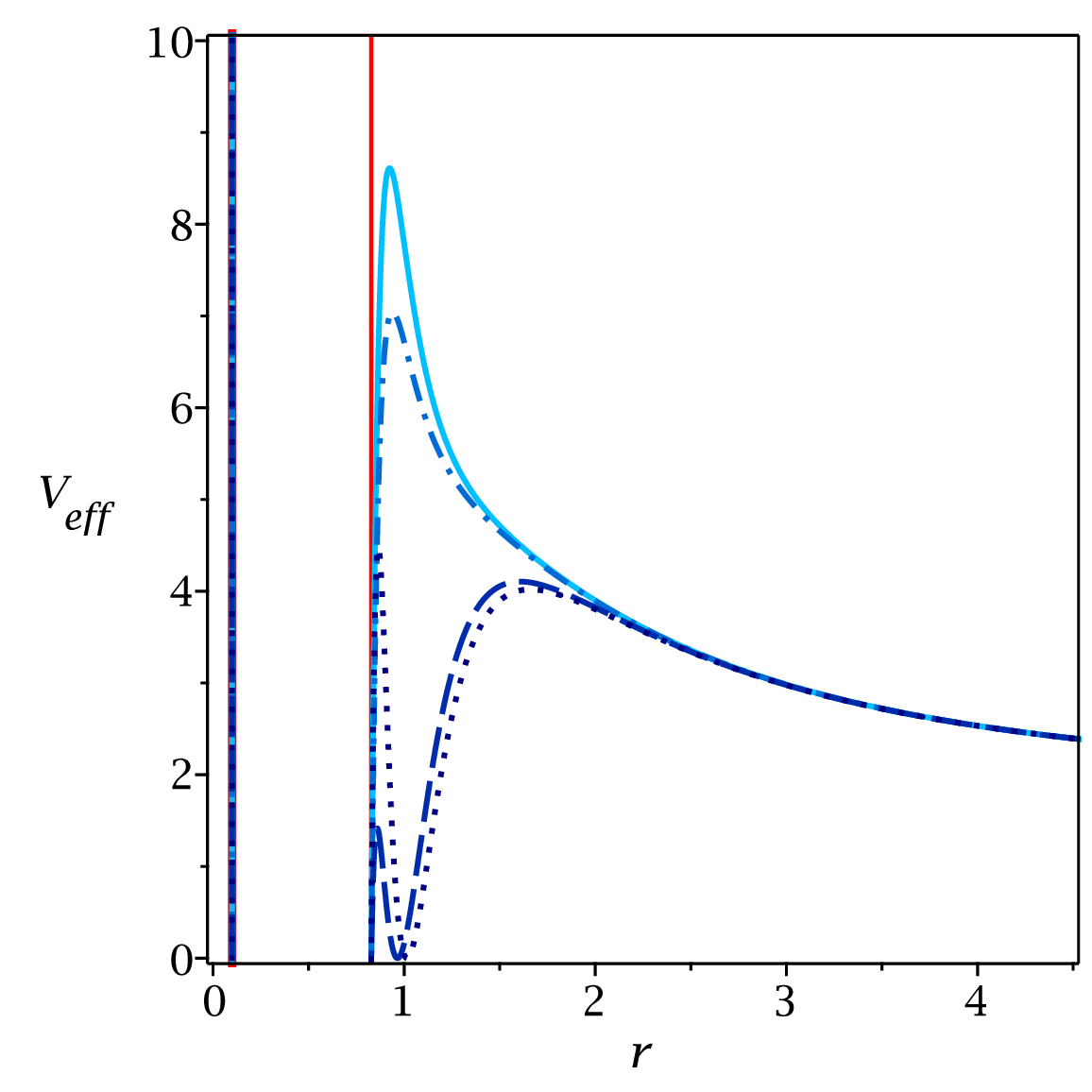}
  }
  \caption{Effective potential \eqref{eqn:Vk1} for test particles moving around a charged black hole with $V(\phi)=\Lambda$ and $k=1$, $m=1$, $L=4$, $q=0.3$, $\Lambda=\frac{1}{3}$. The blue curves represent the effective potential for different values of $b_0$ and the red vertical lines indicate the position of the horizons. Figure (a) is the effective potential for massive particles and figure (b) is the effective potential for massless particles.}
  \label{pic:pot_charged_VLam}
\end{figure}

Here we consider charged spherical black holes with
$V(\phi)=-2\Lambda$. In this case there are two horizons $r_+$ and
$r_-$. We insert the metric functions \eqref{frch2} and
\eqref{grchexp} in the case $k=1$ into the equation of motion
\eqref{eqn:rphi} and the effective potential \eqref{eqn:Veff}
\begin{align}
 \left( \frac{\dd r}{\dd \varphi} \right) ^2 &= \frac{E^2 r^4}{L^2} \left[ 1+ \frac{C_0}{\Lambda^{5/2} r^{5}}
\left( \frac{9}{10}-\frac{\Lambda r^2}{3}-\frac{3m}{4r}+\frac{9
q^2}{14 r^2}-\frac{135}{56 \Lambda r^2} \right)\right]^{-2} \nonumber\\
  &-r^2 \left( 1-\frac{m}{r}+\frac{q^2}{r^2}+\frac{\Lambda  r^2}{3} \right) \left( \frac{\delta r^2}{L^2} -1\right) \, , \\
 V_{\rm eff} &= \left(\delta+\frac{L^2}{r^2}\right) \left( 1-\frac{m}{r}+\frac{q^2}{r^2}+\frac{\Lambda  r^2}{3} \right) \left[ 1+ \frac{C_0}{\Lambda^{5/2} r^{5}}
\left( \frac{9}{10}-\frac{\Lambda r^2}{3}-\frac{3m}{4r}+\frac{9
q^2}{14 r^2}-\frac{135}{56 \Lambda r^2} \right)\right]^2 \, .
\end{align}
Note that we used the expanded version of the function $g(r)$ in
the large $r$ limit. For $r\rightarrow\infty$ the effective
potential diverges for massive particles and approaches $V_{\rm
eff}=\frac{\Lambda L^2}{3}$ for massless particles. At $r=0$ there
is a potential barrier due to the charge. Figure
\ref{pic:pot_charged_VLam} shows the effective potential for
massive particles and light moving around a charged black hole
with $V(\phi)=\Lambda$.

For massive particles we find the following orbit configurations:
\begin{enumerate}
 \item There are two turning points $r_1<r_-$, $r_2>r_+$ and  $\left( \frac{\dd r}{\dd \varphi} \right) ^2\geq 0$ for $r_1 \leq r \leq r_2$. The corresponding orbit is a many-world bound orbit which emerges into another universe each time both horizons are crossed twice.
 \item There are four turning points $r_1<r_-$, $r_+<r_2<r_3<r_4$ and $\left( \frac{\dd r}{\dd \varphi} \right) ^2\geq 0$ for $r_1 \leq r \leq r_2$ and $r_3 \leq r \leq r_4$. A many-world bound orbit and a normal bound orbit exist.
 \item There are six turning points  $r_1<r_-$, $r_+<r_2<r_3<r_4<r_5<r_6$ and $\left( \frac{\dd r}{\dd \varphi} \right) ^2\geq 0$ for $r_1 \leq r \leq r_2$ and $r_3 \leq r \leq r_4$ and $r_5 \leq r \leq r_6$. A many-world bound orbit and two normal bound orbits exist. This configuration does not exist in the Reissner-Nordstr\"om AdS spacetime.
\end{enumerate}
For massless particles we find the following orbit configurations:
\begin{enumerate}
 \item There is a single turning point $r_1<r_-$ and  $\left( \frac{\dd r}{\dd \varphi} \right) ^2\geq 0$ for $r\geq r_1$. The corresponding orbit is a two-world escape orbit which emerges into another universe afther crossing both horizons twice.
 \item There are three turning points $r_1<r_-$, $r_+<r_2<r_3$ and $\left( \frac{\dd r}{\dd \varphi} \right) ^2\geq 0$ for $r_1 \leq r \leq r_2$ and $r \geq r_3$. A many-world bound orbit and an escape bound orbit exist.
 \item There are five turning points  $r_1<r_-$, $r_+<r_2<r_3<r_4<r_5$ and $\left( \frac{\dd r}{\dd \varphi} \right) ^2\geq 0$ for $r_1 \leq r \leq r_2$ and $r_3 \leq r \leq r_4$ and $r \geq r_5$. A many-world bound orbit, a normal bound orbit and an escape orbit exist. This configuration does not exist in the Reissner-Nordstr\"om AdS spacetime.
\end{enumerate}

If the parameter $|C_0|$ is sufficiently small, the geodesics
around the black hole in mimetic gravity behave similar to those
in the Reissner-Nordstr\"om AdS spacetime. As the parameter
$|C_0|$ grows, the effective potential is no longer similar to
Reissner-Nordstr\"om AdS and a minimum at $E=0$ appears in the
effective potential. Then there are stable circular orbits with
zero energy. Moreover, massless particles can move on a stable
bound orbit, which is not possible in the Reissner-Nordstr\"om AdS
spacetime. Note that the behaviour of the effective potential
close to the black hole varies a lot with $C_0$, while at large
$r$ the parameter $C_0$ does not have much influence. For a
description of the effects of the size of the photon sphere see
previous section.

\section{Closing remarks}\label{conc}
In \cite{Der}, Deruelle and Rua showed that, in general, Einstein
field equations of General Relativity are invariant under
disformations, however, relaxing this invariancy can yield the
field equations of mimetic gravity which bring rich physics. In
this theory, in contrast to scalar tensor theories, the mimetic
field $\phi$ is not dynamical by itself and is always restricted
by Eq. (\ref{cond}). Instead, it induces an extra longitudinal
degree of freedom to the gravitation field, in addition to two
transverse degrees of freedom describing gravitons. This extra
degree of freedom, on the cosmological background, admits an
energy density, with geometrical origin, which scales like
pressureless matter and thus mimics dark matter \cite{Mim1}. If
mimetic energy density resembles dark matter, is it possible to
explain the flat galactic rotation curves in this gravity? To
answer this question, in this paper, we have studied topological black
holes in the context of mimetic gravity. We have explored
black hole with various horizon topology including
spherical with positive constant curvature, cylindrical with zero
curvature and hyperbola with negative constant curvature. In order
to reflect the impact of the mimetic field into the spacetime
metric, we should require $\textbf{g}_{tt}\neq-
\textbf{g}_{rr}^{-1}$, which means that we allow an extra degree
of freedom in the metric line elements.

{We have discussed several cases including whether there is or not
a constant potential in the background and whether there is or not
charge on the black holes. We have explored the casual structure
and some physical properties of the solutions. In the absence of
potential, these solutions are not asymptotically flat which is
due to the presence of the mimetic field $\phi$. When a constant
potential (cosmological constant) is taken into account, it
dramatically affects the behaviour of the spacetime and leads to
asymptotically AdS and \textit{approximately} AdS for,
respectively, charged and uncharged solutions. Interestingly
enough, we have noted that when the horizon is spherical, and in
the absence of potential, the spacetime describing by our solution
can explain the flat rotation curves of spiral galaxy without
invoking particle dark matter. In this viewpoint, the dark matter
is indeed a geometrical effect which originate from the extra
degree of freedom of gravitational field, induced into the
spacetime metric, by the mimetic field.} We have discussed the
origin of such behaviour of the metric in ample details. In order
to quantify our results, we applied our metric to a typical spiral
galaxy by assuming that the underlying theory describing the
spacetime is mimetic gravity. We have plotted the orbital speed of
a test particle, at far distance, at galaxies outskirt, around a
typical galaxy with mass $M=10^{12} M_{\odot}$, in terms of the
distance which are in good agreement with observational data
\cite{Man,Bri}.

We have also analyzed the geodesics motion of massive and massless
particles around spherically symmetric black holes in mimetic
gravity. We have focused on two cases: the uncharged case with
zero potential and the charged case with constant potential,
$V(\phi)=-2\Lambda$. We have found that if the parameters $|b_0|$
or $|C_0|$ are sufficiently small, then the geodesics close to the
black hole behave similar to geodesics in Einstein gravity.
However, in the uncharged case at large distances the effective
potential differs from Einstein gravity. Here the logarithmic term
dominates so that massive particles cannot escape the black hole
and will always move on bound orbits. If the parameters $|b_0|$ or
$|C_0|$ grow, then the motion is no longer similar to geodesics in
the Schwarzschild or Reissner-Nordstr\"om (AdS) spacetime. We
observed that, in mimetic gravity, new orbit configurations appear
and there is a minimum with $E=0$ in the effective potential, so
that stable circular orbits with $E=0$ exist. Furthermore, in the
spacetime of a mimetic black hole we found stable bound orbits for
massless particles, which do not exist in Einstein gravity (in
four dimensions). Additionally we found that the size of the
photon sphere is smaller than in the Schwarzschild or
Reissner-Nordstr\"om case for negative  $|b_0|$ or $|C_0|$ and
larger for positive $|b_0|$ or $|C_0|$. However, if $|b_0|$ or
$|C_0|$ get ``too large'' that suddenly the photon sphere shrinks
to the size of the event horizon.

The explanation of the flat galactic rotation curves, in the
background of static spacetime, further supports the viability of
mimetic theory of gravity. It confirms that this theory may serve
an alternative explanation for the presence of dark matter in the
Universe, as pointed out in \cite{Mim1}. Indeed, this
investigation is of great importance because, at least, at
theoretical level, is an indication that the geometrical origin of
dark matter can be understood naturally within a more
comprehensive relativistic theory of gravity. Given the wide
ranges of the observational data available, in the future, we
expect to further constrain our model parameter space and check
the viability of mimetic gravity.

{Finally, we would like to stress that we have only considered zero and constant potential for the mimetic field.
The reason is that only in these cases, we are able to find consistent topological black hole solutions in mimetic gravity.
Finding topological solutions in the presence of variable
potential is not a trivial task and indeed it is out of the scope
of the present work. Even, for uncharged black holes with constant
potential, $V=-2 \Lambda$, it is not easy to find an analytical form for the
metric functions  $g(r)$ (see Eq. (\ref{gr2un})). For example, for
uncharged black holes, if we take the simple form
$V(\phi)={\alpha}/{\phi^2}$ for the potential, then the solution
for $f(r)$  will become
\begin{equation}
f(r)=k-\frac{m}{r}-\frac{\alpha}{2r}
\int{\frac{r^2}{\phi^2(r)}dr}.
\end{equation}
In this case, it is almost impossible to solve the field equations
analytically and find a consistent black hole solution which
satisfy all components of the field equations. This is the reason
why most studies on mimetic black holes consider zero or constant
potential. Let us also note that there is no well-known form for
the mimetic potential which may lead to a consistent black hole
solution. For example in \cite{Odin2}, where the authors
have studied a unified description of early and late time
acceleration in the context of mimetic $F(R)$ gravity, the
potential is not given as an input in the action, but it can be
derived as a complicated function of $H(t)$ and $F(R)$. Thus, it seems in the context
of mimetic black holes, one may need to consider the mimetic potential as
an unknown function and try to construct it from the field equations.
This deserves further and accurate
investigation and we leave it for future studies.}

\begin{acknowledgments}
We thank Jutta Kunz for useful discussions and valuable comments.
AS thanks Shiraz University and also the Institute of Physics of
the University of Oldenburg, for hospitality. S.G. gratefully acknowledges support by the DFG (Deutsche Forschungsgemeinschaft/ German Research Foundation) within the Research Training Group 1620 ``Models of Gravity.''
\end{acknowledgments}

\end{document}